\pdfoutput=1
\documentclass[usenatbib]{mn2e}
\usepackage{apjfonts}
\usepackage{amssymb}
\usepackage{amsmath}
\usepackage{ctable}
\usepackage{url}
\usepackage{fixltx2e} 

\newcommand{\be}{\begin{equation}}
\newcommand{\ee}{\end{equation}}

\newcommand{\etal}{et al.}

\newcommand{\msun}{M_{\sun}}

\newcommand{\paperone}{Paper {\small I}}
\newcommand{\papertwo}{Paper {\small II}}
\newcommand{\msunyr}{M_{\sun}\,{\rm yr^{-1}}}
\newcommand{\tauavg}{\tau_{0}}
\newcommand{\movieurl}{\url{https://www.cfa.harvard.edu/~phopkins/Site/Research.html}}

\newcommand\plotonesize[2]
 {\centering \leavevmode \includegraphics[width={#2\columnwidth}]{#1}}
\newcommand\plotone[1]
 {\centering \leavevmode \includegraphics[width={0.99\columnwidth}]{#1}}

\newcommand{\plotside}[1]
 {\centering \leavevmode \includegraphics[width={0.95\textwidth}]{#1}}

\newcommand{\acknowledgments}{\begin{small}\section*{Acknowledgments}\end{small}}
\newcommand\altaffilmark[1]{$^{#1}$}
\newcommand\altaffiltext[1]{$^{#1}$}
\title[Stellar Feedback and Galactic Winds]{Stellar Feedback in Galaxies and the Origin of Galaxy-scale Winds}

\author[Hopkins \etal]{
\parbox[t]{\textwidth}{ 
Philip F.~Hopkins\thanks{E-mail:phopkins@astro.berkeley.edu}\altaffilmark{1},
Eliot Quataert\altaffilmark{1}, \&
Norman Murray\altaffilmark{2,3} 
}
\vspace*{6pt} \\
\altaffiltext{1}{Department of Astronomy and Theoretical Astrophysics Center, University of California Berkeley, Berkeley, CA 94720} \\
\altaffiltext{2}{Canadian Institute for Theoretical Astrophysics, 
60 St.\ George Street, University of Toronto, ON M5S 3H8, Canada} \\
\altaffiltext{3}{Canada Research Chair in Astrophysics} 
\vspace{-0.6cm}
}

\date{Submitted to MNRAS, September, 2011\vspace{-0.3cm}}
\begin{document}
\maketitle
\label{firstpage}

\begin{abstract}

\vspace{-0.3cm}
Feedback from massive  stars is believed to play a critical role in driving galactic super-winds that enrich the intergalactic medium and shape the galaxy mass function, mass-metallicity relation, and other global galaxy properties.  In previous papers, we have introduced new numerical methods for implementing stellar feedback on sub-GMC through galactic scales in numerical simulations of galaxies;  the key physical processes include radiation pressure in the UV through IR, supernovae (Type-I \&\ II), stellar winds (``fast'' O star through ``slow'' AGB winds), and HII photoionization.  Here, we show that these feedback mechanisms drive galactic winds with outflow rates as high as $\sim10-20$ times the galaxy star formation rate. The mass-loading efficiency (wind mass loss rate divided by the star formation rate) scales roughly as $\dot{M}_{\rm wind}/\dot{M}_{\ast} \propto V_{c}^{-1}$ (where $V_c$ is the galaxy circular velocity), consistent with simple momentum-conservation expectations.  We use our suite of simulations to study the relative contribution of each feedback mechanism to the generation of galactic winds in a range of galaxy models, from SMC-like dwarfs and
Milky-way analogues to $z \sim 2$ clumpy disks.  In massive, gas-rich systems (local starbursts and high-$z$ galaxies), radiation pressure dominates the wind generation.   By contrast, for MW-like spirals and dwarf galaxies the gas densities are much lower and sources of shock-heated gas such as supernovae and stellar winds dominate the production of large-scale outflows. 
In all of our models, however, the winds have a complex multi-phase structure that depends on the interaction between multiple feedback mechanisms operating on
different spatial and time scales: any {\em single} feedback mechanism
fails to reproduce the winds observed.  We use our simulations to provide fitting functions to the wind mass-loading and velocities as a function of galaxy properties, for use in cosmological simulations and semi-analytic models. These differ from typically-adopted formulae with an explicit dependence on the gas surface density that can be very important in both low-density dwarf galaxies and high-density gas-rich galaxies.

\end{abstract}


\vspace{-0.6cm}
\section{Introduction}
\label{sec:intro}

Feedback from massive stars is critical to the evolution of galaxies. 
In cosmological models of galaxy evolution without strong stellar feedback, gas rapidly cools and turns into stars, leading to galaxies with star formation rates much higher than observed, and $\sim$ ten times the stellar mass found in real galaxies \citep[e.g.][and references therein]{katz:treesph,somerville99:sam,cole:durham.sam.initial,
springel:lcdm.sfh,keres:fb.constraints.from.cosmo.sims}. Simply suppressing the rate of star formation does not solve the problem:  the amount of baryons in real galactic disks is much lower than the amount of cool gas in disks found in cosmological simulations, especially in low-mass galaxies
(\citealt{white:1991.galform}; for a recent review see \citealt{keres:2009.gal.mfs.nofb}).
Constraints from the mass-metallicity relation and enrichment of the IGM also imply that the baryons cannot simply be prevented from entering galaxy halos along with dark matter 
\citep{tremonti:mass.metallicity.relation,erb:lbg.metallicity-winds,
aguirre:2001.igm.metal.evol.sims,pettini:2003.igm.metal.evol,songaila:2005.igm.metal.evol}. 
Some process must very efficiently remove baryons from galaxies.

Related problems appear on smaller spatial and time scales. The Kennicutt-Schmidt (KS) law implies that star formation is very slow within galaxies, with a gas consumption time of $\sim50$ dynamical times \citep{kennicutt98}.   Moreover, the integrated fraction of mass turned into stars in GMCs over their lifetime is only a few to several percent  \citep{zuckerman:1974.gmc.constraints,williams:1997.gmc.prop,evans:1999.sf.gmc.review,evans:2009.sf.efficiencies.lifetimes}.  Without strong stellar feedback, however, self-gravitating collapse leads to most of the gas turning into stars  in just a few dynamical times. 

The problem, then, on both galactic and sub-galactic scales, is twofold. First, star formation must be ``slowed down'' at a given global/local gas surface density.  But this alone would still violate  integral constraints,  producing galaxies and star clusters more massive than observed  by an order of magnitude. 
Thus the second problem: on small scales gas must be expelled from GMCs, and on galactic scales either prevented from entering, or, more likely in our judgment, removed from, the host galaxy. In other words, local outflows  and global super-winds must be generated that can remove gas at a rate rapid compared to the star formation rate. 

Because low-mass galaxies are preferentially baryon-poor, 
matching the faint end of the observed galaxy mass function in cosmological simulations requires that the global efficiency of galactic super-winds scales as a declining power of galaxy mass or circular velocity.
\citet{oppenheimer:outflow.enrichment} and \citet{oppenheimer:recycled.wind.accretion} 
find that an average scaling $\dot{M}_{\rm wind}/\dot{M}_{\ast}\propto V_{c}^{-1}$ 
produces reasonably good agreement with the observed mass functions at different redshifts,\footnote{Note that 
the agreement we speak of is with the faint sub-$L_{\ast}$ end of the mass function. It is widely agreed that different physics, 
perhaps AGN feedback, is critical for the regulation of the bright super-$L_{\ast}$ end of the mass function. 
We focus here on stellar feedback and star-forming systems.} with a normalization 
such that an SMC-mass dwarf has a mass loading $\dot{M}_{\rm wind}/\dot{M}_{\ast}\sim10$ (although this scaling may still over-produce the number of very low-mass galaxies). 
Large mass-loading factors of several times the SFR are also estimated  in even relatively 
massive local galaxies and massive star-forming regions at $z\sim2-3$ 
\citep{martin99:outflow.vs.m,
martin06:outflow.extend.origin,heckman:superwind.abs.kinematics,newman:z2.clump.winds.prep,
sato:2009.ulirg.outflows,chen:2010.local.outflow.properties,
steidel:2010.outflow.kinematics,coil:2011.postsb.winds}.

The scaling $\dot{M}_{\rm wind}/\dot{M}_{\ast}\propto V_{c}^{-1}$ is expected from momentum conservation arguments, given a number of simplifying assumptions and sufficient global momentum input from supernovae, stellar winds, radiation pressure, etc.  \citep{murray:momentum.winds}. 
Direct observations, while uncertain, tend to favor velocity and $\dot M_{\rm wind}$ scalings similar to this constraint for the bulk of the outflowing gas \citep{martin05:outflows.in.ulirgs,rupke:outflows,
weiner:z1.outflows}.

To date, however, numerical simulations have generally not been able to {produce}, from 
an a priori model, winds with either such large absolute mass loading 
factors or the  scaling of mass-loading with galaxy mass/velocity. 
Many simulations, lacking the ability to directly resolve the relevant feedback 
processes,  put in winds ``by hand'' by e.g. forcing an outflow rate that scales in a user-specified manner with the star formation rate or other parameters 
\citep{springel:multiphase,
oppenheimer:metal.enrichment.momentum.winds,
sales:2010.cosmo.disks.w.fb,genel10}. 
Alternatively, models that  self-consistently include stellar feedback have generally 
been limited to a small subset of the relevant processes; the vast majority include only thermal feedback via supernovae (i.e.\ thermal 
energy injection with some average rate that scales with the mass in young stars). 
However, thermal feedback is very inefficient in the dense regions where star formation occurs, 
and in the ISM more broadly in gas-rich galaxies. 
For this reason, such models require further changes to the physics in order for thermal energy injection to have a significant effect.  Often cooling (along with star formation and other 
hydrodynamic processes) is ``turned off'' for an extended period of time 
\citep{thackercouchman00,governato:disk.formation,brook:2010.low.ang.mom.outflows}. 
With or without these adjustments, however, such models  generally obtain winds that are  weaker than those required to explain the galaxy mass function, especially at low masses \citep[see e.g.][and references therein]{guo:2010.hod.constraints,powell:2010.sne.fb.weak.winds,brook:2010.low.ang.mom.outflows,nagamine:2010.dwarf.gal.cosmo.review}.   

In our view, part of the resolution of this difficulty lies in the treatment of the ISM physics {\em within} galaxies.   Feedback processes other than supernovae are critical for suppressing star formation in dense gas; these include protostellar jets, HII regions, stellar winds, and radiation pressure from young stars.  Including  these mechanisms self-consistently maintains a reasonable fraction of the ISM at densities where the thermal heating from supernovae has a larger effect.  This conclusion implies that (not surprisingly) a realistic treatment of galactic winds requires a more realistic treatment of the stellar feedback processes that maintain the multi-phase structure of the ISM of galaxies.

Motivated by this perspective, in \citet{hopkins:rad.pressure.sf.fb} (\paperone) and \citet{hopkins:fb.ism.prop} (\papertwo) 
we developed a new set of numerical models to follow feedback on small scales in GMCs and star-forming regions, in simulations with pc-scale resolution.\footnote{\label{foot:url}Movies of these
  simulations are available at \movieurl} 
These simulations include the momentum imparted locally (on sub-GMC 
scales) from stellar radiation pressure, radiation pressure on larger scales via the light that escapes star-forming regions, HII photoionization heating, as well as the heating, momentum deposition, and mass loss by SNe (Type-I and Type-II)  and stellar winds (O star and AGB).  The feedback is tied to the young stars, with the energetics and time-dependence 
taken directly from stellar evolution models.     Our models also include realistic cooling to temperatures $<100\,$K, and a treatment of the molecular/atomic transition in gas and its effect on star formation \citep[following][]{krumholz:2011.molecular.prescription}.

We showed in Papers I \& II that these feedback mechanisms produce a quasi-steady ISM  in which giant molecular clouds form and disperse 
rapidly, after turning just a few percent of their mass into stars.   This leads to an ISM with phase structure, turbulent velocity dispersions, scale heights, and GMC properties (mass functions, sizes, scaling laws) in reasonable agreement with observations.   In this paper, we use these same models of stellar feedback to quantitatively {\em predict} the elusive winds invoked in almost all galaxy formation models.   

The remainder of this paper is organized as follows. In \S \ref{sec:sims} we summarize 
the galaxy models we use and our methods of implementing stellar feedback.  In \S~\ref{sec:morph} we discuss how each feedback mechanism 
affects the morphology, phase structure, and 
velocity distribution of galactic winds.   In \S~\ref{sec:winds}, we discuss the mass-loading of winds and how the  mass loss rate depends on 
the inclusion of  different feedback mechanisms. We further determine how the outflow rate scales with galaxy properties and use our simulations to derive more accurate 
approximations to wind scalings for use in cosmological simulations and semi-analytic models.
In \S~\ref{sec:discussion} we summarize our results and discuss their implications.  

\vspace{-0.3cm}
\section{Methods}
\label{sec:sims}

The simulations used here are described in detail in 
\paperone\ (see their Section~2 and Tables~1-3) and \papertwo\ (their Section~2).
However we briefly summarize the most important properties of the models here. 
The simulations were performed with the parallel TreeSPH code {\small 
GADGET-3} \citep{springel:gadget}.  They include stars, dark matter, and gas, 
with cooling, star formation, and stellar feedback.

Gas follows a standard atomic cooling curve but in addition can cool to 
$<100\,$K via fine-structure cooling. This allows it to collapse to very high 
densities, and star formation occurs in dense regions above a threshold 
$n>1000\,{\rm cm^{-3}}$, with a rate $\dot{\rho}_{\ast}=\epsilon\,\rho/t_{\rm ff}$ 
where $t_{\rm ff}$ is the free-fall time and $\epsilon=1.5\%$ is an efficiency 
taken from observations of star-forming regions with the same densities 
\citep[][and references therein]{krumholz:sf.eff.in.clouds}. 
We further follow \citet{krumholz:2011.molecular.prescription} and calculate the molecular 
fraction within the dense gas as a function of the local column density and 
metallicity, and allow star formation only from that gas. 
In \paperone\ we show that the SFR is essentially independent of the small-scale star formation law (robust to large variations in the 
threshold and efficiency, and changes in the power-law index 
$\dot{\rho}\propto \rho^{1-2}$).   This is because star formation is feedback-regulated and dense star-forming regions grow in mass until sufficient new stars have formed
to halt further collapse. Likewise in \papertwo\ we show 
that the corrections from the molecular chemistry are negligible 
at the masses and metallicities we model.

\vspace{-0.5cm}
\subsection{Disk Models}

Our calculations span four distinct initial disk models, designed to 
represent a range of characteristic galaxy types. Each initial disk has a 
bulge, stellar and gaseous disk, and dark matter halo.  The disks are initialized in equilibrium so that in the absence of cooling, star formation, and 
feedback there are no significant transients. The gaseous disk is initially 
vertically pressure-supported, but this thermal energy is radiated away in much less than a dynamical  time and the emergent vertical structure depends on feedback. Our 
``low'' resolution runs (used to evolve the simulations for 
several Gyr, to ensure steady-state behavior) use $\approx3\times10^{6}$ particles, with 
$\approx10^{6}$ particles in the disk, giving SPH smoothing 
lengths of $\sim10$\,pc in the central few kpc of a MW-like disk (the smoothing length scales 
linearly with the disk size/mass scale). 
Our ``standard'' resolution cases use $\sim30$ times as many particles, 
and correspondingly have $\sim1-5\,$pc smoothing lengths 
and particle masses of $500\,\msun$; these are run for a few orbital times each. A few ultra-high resolution runs used for convergence tests employ 
$\sim10^{9}$ particles, with sub-pc resolution on kpc scales.

(1) SMC: an SMC-like dwarf, with baryonic mass $M_{\rm bar}=8.9\times10^{8}\,\msun$ 
and halo mass $M_{\rm halo}=2\times10^{10}\,\msun$ (concentration $c=15$), 
a \citet{hernquist:profile} profile bulge with a mass $m_{b}=10^{7}\,\msun$, and exponential 
stellar ($m_{d}=1.3\times10^{8}\,\msun$) and gas disks ($m_{g}=7.5\times10^{8}\,\msun$) 
with scale-lengths $h_{d}=0.7$ and $h_{g}=2.1$\,kpc, respectively. 
The initial stellar scale-height is $z_{0}=140$\,pc and the stellar disk is initialized such that the 
Toomre $Q=1$ everywhere. The gas and stars are initialized with uniform metallicity $Z=0.1\,Z_{\sun}$.

(2) MW: a MW-like galaxy, with halo and baryonic properties of $(M_{\rm halo},\,c)=(1.6\times10^{12}\,\msun,\,12)$ and $(M_{\rm bar},\,m_{b},\,m_{d},\,m_{g})=(7.1,\,1.5,\,4.7,\,0.9)\times10^{10}\,\msun$, $Z=Z_{\sun}$, and scale-lengths $(h_{d},\,h_{g},\,z_{0})=(3.0,\,6.0,\,0.3)\,{\rm kpc}$. 

(3) Sbc: a LIRG-like galaxy (i.e.\ a more gas-rich spiral than is characteristic 
of those observed at low redshifts)
with $(M_{\rm halo},\,c)=(1.5\times10^{11}\,\msun,\,11)$,  
$(M_{\rm bar},\,m_{b},\,m_{d},\,m_{g})=(10.5,\,1.0,\,4.0,\,5.5)\times10^{9}\,\msun$, 
$Z=0.3\,Z_{\sun}$, 
and $(h_{d},\,h_{g},\,z_{0})=(1.3,\,2.6,\,0.13)\,{\rm kpc}$. 

(4) HiZ: a high-redshift massive starburst disk, chosen to match the 
properties of the observed non-merging but rapidly star-forming SMG 
population, with 
$(M_{\rm halo},\,c)=(1.4\times10^{12}\,\msun,\,3.5)$ and a 
virial radius appropriately rescaled for a halo at $z=2$ rather than $z=0$,  $(M_{\rm bar},\,m_{b},\,m_{d},\,m_{g})=(10.7,\,0.7,\,3,\,7)\times10^{10}\,\msun$, $Z=0.5\,Z_{\sun}$,  
and 
$(h_{d},\,h_{g},\,z_{0})=(1.6,\,3.2,\,0.32)\,{\rm kpc}$. 

\vspace{-0.5cm}
\subsection{Feedback Models}

The most important physics in these simulations is the model of 
stellar feedback.   
We include feedback from a variety of mechanisms, each of which we briefly describe below.  More details about our implementations of this physics are given in \paperone\ and \papertwo.
 We use a  \citet{kroupa:imf} initial mass function (IMF) throughout and use STARBURST99 \citep{starburst99} to calculate the stellar luminosity, mass return from stellar winds, supernova rate, etc. as a function of the age and metallicity of each star particle.

(1) {\bf Local Momentum Deposition} from Radiation Pressure, 
Supernovae, \&\ Stellar Winds: In \paperone, we present the radiation pressure aspect of this 
model for feedback from young star clusters in detail. At each timestep, gas particles identify the nearest density peak representing the center of the nearest star-forming ``clump'' or GMC-analog. 
We calculate the total luminosity of the star particles inside the sphere defined by the distance from the center of this star-forming region to the gas particle of interest;  the incident flux on the gas is then determined assuming that the local star forming region is optically thick to the UV radiation.

The rate of momentum deposition from radiation pressure is then $\dot{P}_{\rm rad}\approx (1+\tau_{\rm IR})\,L_{\rm incident}/c$ 
where the term 
$1+\tau_{\rm IR}$ accounts for the fact that most of the initial optical/UV radiation is 
absorbed and re-radiated in the IR; $\tau_{\rm IR}=\Sigma_{\rm gas}\,\kappa_{\rm IR}$ 
is the optical depth in the IR, which allows for the fact that the momentum is boosted 
by multiple scatterings in optically thick regions. 
Here $\Sigma_{\rm gas}$ is calculated self-consistently as the average surface density of the identified clump, with $\kappa_{\rm IR}\approx5\,(Z/Z_{\sun})\,{\rm g^{-1}\,cm^{2}}$ approximately 
constant over the relevant physical range of dust temperatures. 
The imparted acceleration is directed along the flux vector.   In \paperone\ we discuss numerous technical aspects of this implementation -- such as the effects of resolution, photon leakage, 
and how the momentum is discretized -- and show that essentially all our conclusions are robust to uncertainties in these choices. 

The {\em direct} momentum of SNe ejecta and stellar winds $\dot{P}_{\rm SNe}$ 
and $\dot{P}_{\rm w}$ are similarly tabulated from STARBURST99 and injected as an appropriate function of age and metallicity to the gas within a smoothing length of each star. This source of turbulent energy is almost always smaller than that due to radiation pressure discussed above.  In some cases, however, in particular in dwarf galaxies, the work done by bubbles of gas shock-{\em heated}  by supernovae and/or stellar winds is dynamically important; this is discussed below. 

(2) {\bf Supernova and Stellar Wind Shock-Heating}: The gas shocked by 
supernovae and stellar winds can be heated to high temperatures, generating bubbles and 
filaments of hot gas. We tabulate the Type-I and Type-II SNe rates from \citet{mannucci:2006.snIa.rates}  and STARBURST99, respectively, as a function of age and 
metallicity for all star particles, and stochastically determine at 
each timestep if a SNe occurs.  That is, the SNe are resolved 
discretely in time (as opposed to continuous energy injection). For each SNe, the appropriate thermal energy is injected into the gas within a smoothing length of the star particle. Similarly, stellar winds are assumed to shock locally and so we inject the appropriate tabulated mechanical 
power $L(t,\,Z)$ as a continuous function of age and metallicity into the gas within a smoothing length of the star particles.   The specific energy of these stellar winds is large for young stellar populations in which fast winds from massive stars dominate, but declines rapidly at later times when slower, AGB winds dominate. 

(3) {\bf Gas Recycling:} Gas mass is returned continuously
to the ISM from stellar evolution, at a rate tabulated from SNe and stellar mass 
loss in STARBURST99. The integrated mass fraction recycled 
is $\sim0.3$.  

(4) {\bf Photo-Heating of HII Regions}: For each star particle, we tabulate the 
rate of production of ionizing photons from STARBURST99. Starting from the nearest gas particle and working our way outwards radially from the star, we then 
ionize each gas particle which is not  already ionized until the photon budget is exhausted 
(using the gas and stellar properties to determine the photon production rate needed to maintain 
the particle as fully ionized). Gas which is ionized is immediately heated to $\sim10^{4}$ K, unless is already above this temperature; moreover, the gas is not allowed to cool below $10^4$ K until it is no longer in an HII region. This method allows for overlapping, non-spherical HII regions that can extend to radii $\sim$kpc. 

(5) {\bf Long-Range Radiation Pressure:} Radiation pressure from photons absorbed 
in the immediate vicinity of stars is captured in mechanism (1). However, photons that 
escape these regions can still be absorbed at larger radii. For each star particle, we 
construct the intrinsic SED ($L_{\nu}$) as a function of age and metallicity; 
we then use the local density and density gradients to estimate the integrated 
column density and attenuation of the SED using $\tau_{\nu} = \kappa_{\nu}\,\Sigma \approx \kappa_\nu \, \rho\,(h_{\rm sml} + |\nabla\ln\rho|^{-1})$, where $h_{\rm sml}$ is the smoothing length and $\kappa_\nu$ is the frequency-dependent opacity (assuming dust opacities that 
scale with metallicity, as in (1) above).   The resulting ``escaped'' SED gives a frequency-dependent 
flux ${\bf F}_{\nu}$ that is propagated to large distances. We construct a force tree for 
this long range force in an identical fashion to the gravity tree, since after attenuating the flux near the star particle, the stellar flux is assumed to decrease $\propto r^{-2}$. Each gas particle then sees an incident net flux vector  ${\bf F}_{\nu}^{i}$, integrated over all stars in the galaxy. 


Extensive numerical tests of the feedback models are presented in \paperone\ and 
\papertwo. In the Appendix, we present additional numerical and convergence studies 
specific to galactic winds.

\section{Wind Morphologies \&\ Phase Structure}
\label{sec:morph}

\subsection{Morphologies with Feedback} 
\label{sec:morph:all}

\begin{figure*}
    \centering
    \plotside{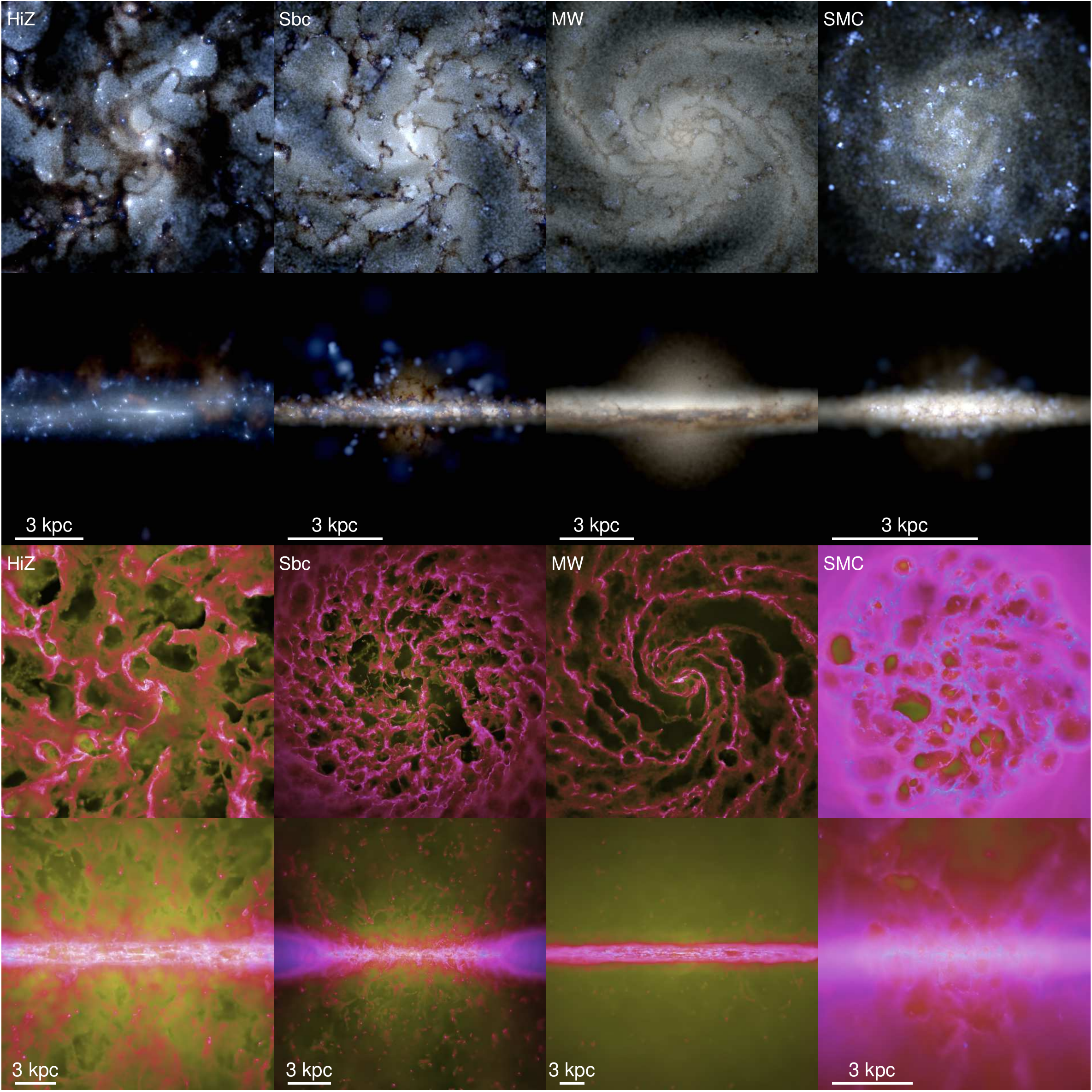}
    \caption{Morphology of the gas \&\ stars in our standard simulations with all feedback mechanisms enabled; we show face-on and edge-on projections.   The time in each is $\sim2$ orbital times after the simulation begins, when the disk is in a feedback-regulated steady state. 
    {\em Top:} Stars. We show a mock $ugr$ (SDSS-band) composite, 
    with stellar spectra calculated from their known ages and metallicities 
    and including dust attenuation.
    Brightness is logarithmically scaled with a $\approx2\,$dex stretch.
    {\em Bottom:} Gas. Brightness encodes projected gas density (logarithmically 
    scaled with a $\approx4\,$dex stretch); color encodes gas temperature 
    with blue being $T\lesssim1000\,$K molecular gas, pink $\sim10^{4}-10^{5}$\,K 
    warm ionized gas, and yellow $\gtrsim10^{6}\,$K hot gas. 
    Each column corresponds to one of our four galaxy models. {\bf HiZ:} A 
    massive, $z\sim2-4$ starburst disk with $\dot{M}_{\ast}>100\,\msun\,{\rm yr^{-1}}$; 
    gravitational collapse forms kpc-scale complexes and a clumpy morphology, 
    with violent outflows containing dense, cold gas driven by the massive starburst.
    {\bf Sbc:} A $z\sim0$ dwarf starburst galaxy; the disk is clumpy but more contained in 
    global spiral structure, with outflows from the central few kpc producing a multi-phase 
    clumpy wind.     {\bf MW:} A MW-analogue; the gas morphology more closely follows stars in a 
    global spiral pattern; the ``wind'' here is primarily hot SNe-driven bubbles/holes ``venting'' 
    out, rather than the clumpy filaments/streamers seen in the HiZ/Sbc cases.
    {\bf SMC:} An isolated SMC-analogue dwarf; the disk is thick with 
    irregular star formation and large bubbles from overlapping SNe; the wind is prominent 
    and contains a mix of hot gas and entrained cold/warm material in filaments/loops/arcs.
   The ``smooth'' outer disk  is ionized and low-density, sufficient to prevent collapse at these masses. \label{fig:morph.gas}}
\end{figure*}

Figure~\ref{fig:morph.gas} shows the stellar and gas morphology for each galaxy model in our standard simulations with all feedback mechanisms included. All cases exhibit clear outflows.
The stellar maps show a mock $ugr$ composite image in which the spectrum of each 
star particle is calculated from its age and metallicity, and the 
dust extinction/reddening is accounted for from the line-of-sight dust mass 
(following \citealt{hopkins:lifetimes.letter}).
The gas maps show the projected gas density (intensity) 
and temperature (color, with blue representing cold molecular gas at 
$T\lesssim 1000\,$K, pink representing the warm ionized gas at $\sim10^{4}-10^{5}$\,K, 
and yellow representing the hot, X-ray emitting gas at $\gtrsim10^{6}\,$K).  

The HiZ models result in disks with massive ($>10^{8}\,\msun$) 
$\sim$kpc-scale clumps, similar to ``clump-chain'' systems. Violent outflows arise 
from throughout the disk driven by the very high rate of star formation, with $\dot{M}_{\ast}>100\,\msun\,{\rm yr^{-1}}$.  The outflow is clearly multi-phase: there is a volume-filling hot component (largely from hot gas  ``venting'' out of SNe-heated regions and from ``fast'' OB stellar winds), but most of the mass
is in dense clumps, filaments, and streams.  We quantify this more in \S \ref{sec:winds:phases.coldwarmhot}.

The Sbc models also produce clumpy gas structure 
but on smaller scales and with the global structure more closely tracing spiral arms.   This is
similar to observed dwarf starbursts (e.g. NGC 1569 or 1313).  Again, a 
multi-phase super-wind is evident, arising from the reasonably high specific star formation rate 
of $\dot{M}_{\ast}/M_\ast \sim 0.2-1 \ {\rm Gyr^{-1}}$.

In the MW models, which are much more gas poor than the HiZ and Sbc models, the gas in the edge-on image traces the stellar spiral and bar instabilities, albeit with filamentary and clumpy GMC structure internal to the arms. The outflows here are less clearly multi-phase; with the low specific SFR and gas fraction, much of the material is directly ``vented'' hot gas, rather than accelerated cold material. 

The SMC models have a patchy ISM typical of dwarf irregular systems, with large 
overlapping SNe bubbles. Outflows above the thick disk appear as a mix of venting hot gas and 
shells of warm material accelerated by the hot gas; the latter appear visually as loops, shells, and filaments of gas.
 
\subsection{Effects of Each Feedback Mechanism in Turn} 
\label{sec:morph:fbmodel}

\begin{figure*}
    \centering
    \plotside{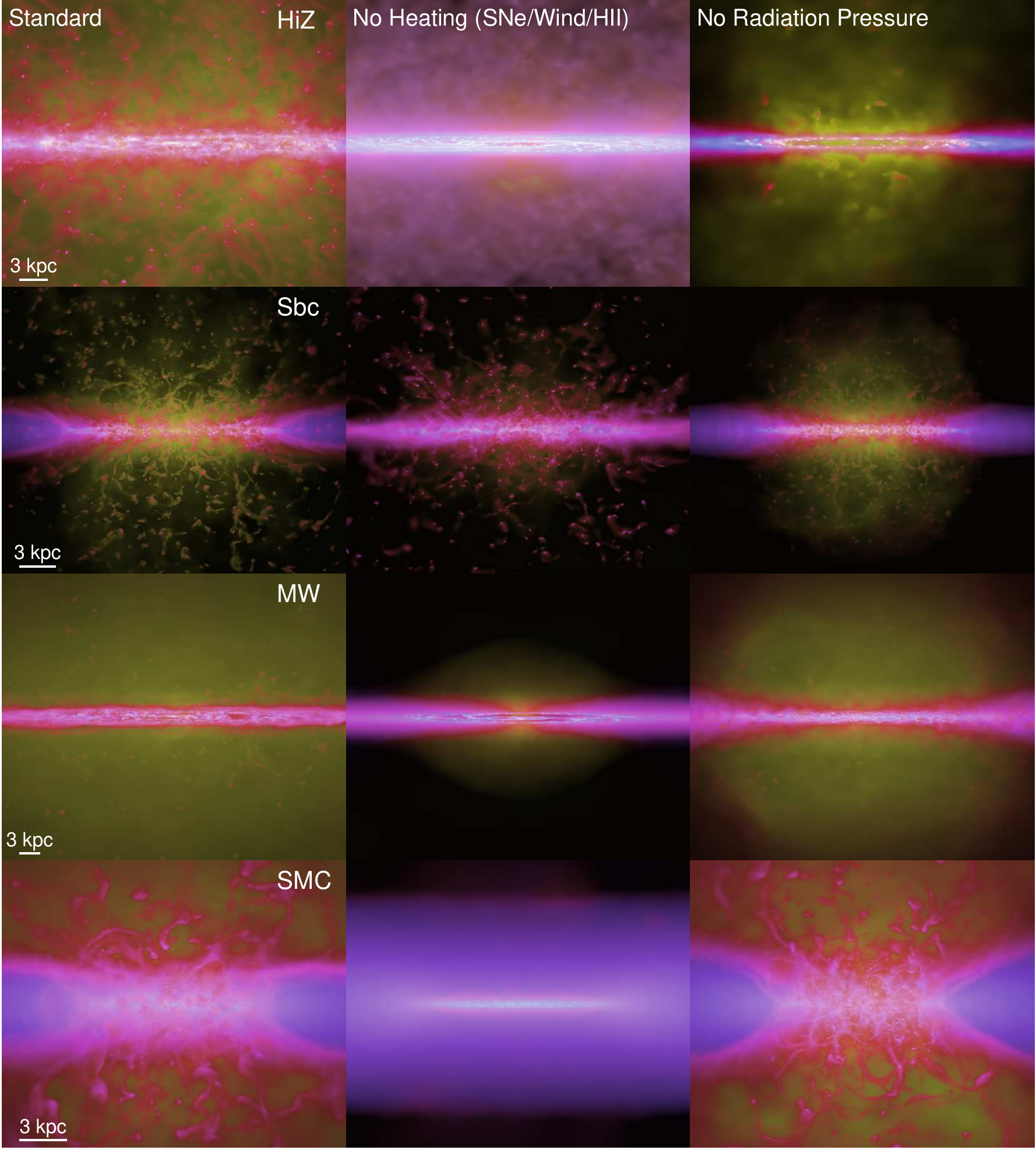}
    \caption{Edge-on gas morphologies (as Fig.~\ref{fig:morph.gas}), with 
    different feedback mechanisms enabled.
    {\em Left:} Standard model (all feedback on).
    {\em Middle:} No heating: energy injection from SNe, shocked stellar 
    winds, and HII photo-heating is disabled (gas recycling from these 
    mechanisms remains, but is ``cold''). The ``hot'' volume-filling ISM is greatly diminished 
    but in the HiZ/Sbc cases the global morphology and much of the wind 
    mass (in the cold/clumpy phase) remains.
    In the MW/SMC cases, removing heating eliminates most of the wind mass 
    (including warm gas, previously accelerated by hot gas).
    {\em Right:} No radiation pressure momentum flux (local or long-range). The 
    hot medium remains, but in the HiZ/Sbc cases most of the cold/warm clumpy 
    material in the wind is absent (hot gas simply vents). The wind in 
    the MW/SMC cases is less strongly affected, but within the disk, GMCs collapse much further than they would with radiation  pressure, giving a higher SFR at the same absolute wind mass. 
    \label{fig:sfhwind.morph.vs.fb.hiz}}
\end{figure*}

Figure~\ref{fig:sfhwind.morph.vs.fb.hiz} shows the edge-on wind morphologies of models  that include different feedback mechanisms, but  are otherwise identical.

First consider the HiZ model.  If we view the disk face-on (see \papertwo), the 
clumpy morphology seen in Figure~\ref{fig:morph.gas} is evident in all cases (this arises from violent gravitational 
instability); however, without momentum from radiation pressure, the 
characteristic sizes of individual clumps collapse from $\sim$kpc to $\sim$pc 
and the gas piles up at densities $\sim10^{6}\,{\rm cm^{-3}}$. 
Supernovae, stellar wind momentum, and HII photoionization pressure are not sufficiently strong to resist the collapse of clumps beyond a critical threshold 
and the cooling time in these dense regions is much too short for thermal pressure to be important. Some hot gas is produced from star clusters that have nearly exhausted their gas supply.  This  vents directly out of the disk producing the hot gas seen in the right-most image in 
Figure~\ref{fig:sfhwind.morph.vs.fb.hiz}, but the warm/cold 
material in the wind -- most of the wind mass -- is almost totally absent.    

If we instead turn off sources of gas heating (SNe, stellar wind shock-heating, and 
HII photo-heating), the primary difference relative to our standard run is 
in the volume-filling diffuse phase (the properties interior to dense clumps are nearly identical). 
The diffuse gas is much cooler (note the absence of yellow gas in the middle panel), but 
the warm clumps in the outflow -- most of the wind mass -- remain, directly accelerated 
out of the disk by radiation pressure.

For the MW-analogue model without radiation pressure, GMCs again collapse to significantly smaller sizes, although the runaway is not as severe because ``fast'' stellar winds and HII heating are able to somewhat suppress runaway GMC collapse (see \papertwo). Some 
wind material in the cold phase is also absent, but this is not the dominant wind phase in the full simulations, so the net difference is much weaker than in the HiZ case. Removing the sources of 
heating has a comparable effect on the SFR as removing the radiation pressure 
(\papertwo), but it removes the volume-filling ``hot'' gas and dramatically suppresses the wind 
(middle panel in Fig.~\ref{fig:sfhwind.morph.vs.fb.hiz}); there is thus little material accelerated  by radiation pressure alone in MW-like models.

Figure~\ref{fig:sfhwind.morph.vs.fb.hiz} shows that the Sbc case lies, as expected, somewhere between the MW and HiZ cases in terms of the contribution of the different feedback processes to the galaxy-scale outflow.    In the SMC-like model, on the other hand, the average gas density is very low ($\lesssim 0.1\,{\rm cm^{-3}}$), approaching the regime in which the cooling time can be comparable to the dynamical time.  In addition, with the lower densities and metallicities the IR optical depths are not large. As a result, with only radiation pressure present, there is 
no hot gas and essentially no wind.   Turning off SNe heating, by contrast, has a large effect -- the volume filling factor of hot gas ``bubbles'' drops, and the wind is basically absent.

\begin{figure}
    \centering
    \plotonesize{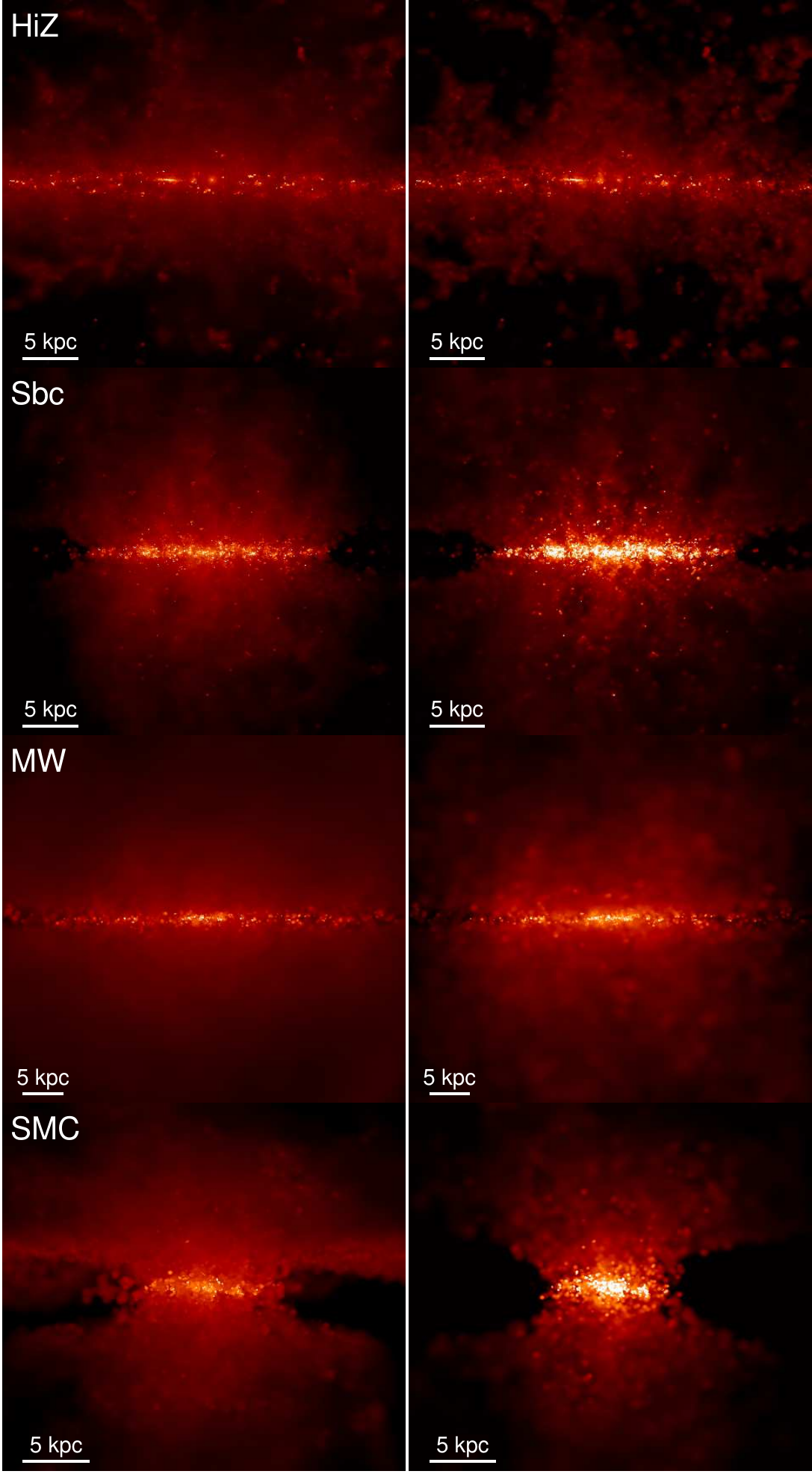}{1.02}
    \caption{{\em Left:} Galactic wind thermal emission morphology, as a proxy for X-ray emission. The maps show 
    edge-on images of our standard models for each disk (with all feedback enabled).
    Intensity here encodes the projected bremsstrahlung emissivity 
    (eq.~\ref{eqn:bremsstrahlung}). A roughly bi-conical, clumpy/multiphase 
    wind is typical, similar to that observed in, e.g., M82. {\em Right:} Same, but now showing the metal cooling luminosity. Note the enhanced clumpiness from the inhomogenous metal distribution.
    \label{fig:wind.phase.morph}}
\end{figure}

\begin{figure*}
    \centering
    \plotside{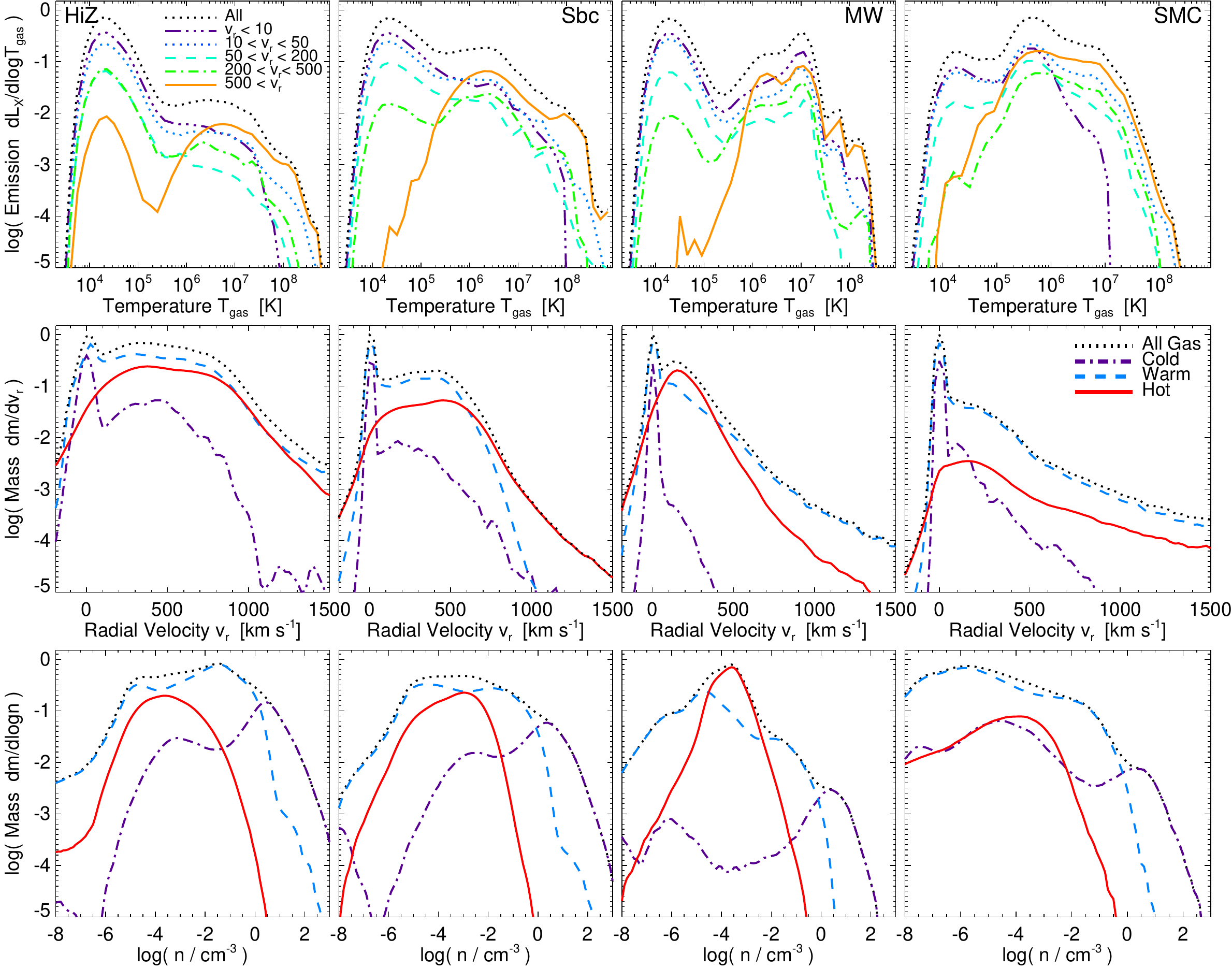}
    \caption{{\em Top:}  Thermal bremsstrahlung emission-weighted distribution of gas temperature
    $dL_{X}/d\log{T_{\rm gas}}$. 
    For each galaxy model (columns), we show results for our standard simulation with all feedback mechanisms 
    (time-averaged over the simulation).
    Different lines in a given panel correspond to gas with different radial velocities (relative to the center of mass). 
    The emission from more rapidly outflowing wind material tends to be 
    dominated by the hotter gas with $T \sim 10^{6-8}$ K. 
    {\em Middle:} Mass-weighted distribution of gas outflow radial velocities $dm/dv_{r}$.  The distribution for all gas is divided into the 
    contribution from different phases: cold (primarily molecular) ($T<2000\,K$), 
    warm (primarily ionized) ($2000<T<4\times10^{5}\,K$) and hot diffuse ($T>4\times10^{5}\,K$) gas.
    The high $v_{r}$ part of the distribution is the wind material, with a distribution  
    that can be fit by equation~\ref{eqn:windmass.vs.vr}. 
    This outflowing gas consists primarily of a mix of warm and hot gas with some colder (likely molecular) material. The warm material usually dominates the mass, 
    and is in the form of the dense clouds, filaments, and shells seen in 
    Figure~\ref{fig:morph.gas}.
    {\em Bottom:} Mass-weighted density distribution for the unbound wind material, 
    divided into different phases as in the middle panel.
    The cold/warm/hot gas dominates at high/intermediate/low densities, respectively. 
    The ``warm'' material at very low densities, $\ll 10^{-4}\,{\rm cm^{-3}}$ 
    is previously ``hot'' material that has adiabatically cooled as it expands; 
    these low densities arise artificially because we do not include an IGM into which the wind expands.
    \label{fig:wind.phases}}
\end{figure*}

\subsection{Appearance in X-Rays}
\label{sec:winds:phases.xr}

X-ray observations of canonical systems such as M82 provide a strong probe of the shock-heated phase of galactic winds.   We therefore consider some of the X-ray properties of our simulated galactic outflows.  For convenience, rather than making a detailed mock observation corresponding to a given instrument, sensitivity, redshift, and energy range, we instead quantify the thermal bremsstrahlung emission, for which the emissivity per unit volume is
\be
\label{eqn:bremsstrahlung}
u_{X} \propto T_{\rm gas}^{1/2}\,n_{e}\,n_{i}
\ee
where $T_{\rm gas}$ is the gas temperature and $n_{e}$ and $n_{i}$ 
are the number density of electrons and ions.   At fixed pressure, $u_X$ is typically dominated by the lowest temperature gas and so equation~\ref{eqn:bremsstrahlung} need not refer specifically to X-ray observations.   We quantify the contributions of gas at different temperatures to the thermal emission below.  To start, however, Figure~\ref{fig:wind.phase.morph} shows the observed morphology of the winds for our fiducial simulation of each galaxy model, with surface brightness weighted (logarithmically) using the thermal emission in equation~\ref{eqn:bremsstrahlung} (using the known temperature, density, and ionization information in the simulations). 
Very broadly, the appearance of the winds in Figure~\ref{fig:wind.phase.morph} is similar to that seen in 
Figure~\ref{fig:morph.gas}, but the clumpy structure of the winds is more evident and
the global morphology is more obviously bi-conical (since much of the in-plane material 
is neutral and so does not appear here). Overall, the morphologies of the Sbc and SMC-like models are qualitatively similar to observations of systems like M82. The HiZ case is more spatially extended, making the outflow less columnar.   The MW model is less clearly wind-like but is instead consistent with a diffuse halo.

The top panel of Figure~\ref{fig:wind.phases} shows  $d L_X/d\ln T_{\rm gas}$, the contribution of gas at different temperatures to the total thermal bremsstrahlung emission. 
This weighting favors dense cool gas given roughly comparable pressures at different temperatures, but there is a rapid cutoff below $\sim 10^{4}\,$K because the gas becomes neutral.   Figure~\ref{fig:wind.phases} shows that much of the "x-ray" emission would be contributed by gas at $\sim10^{6}-10^{7}\,$K with the exact range of temperatures depending on the system. Figure~\ref{fig:wind.phases} also divides $d L_X/d\ln T_{\rm gas}$ into intervals in radial outflow velocity. The material at low $v_{r}<10\,{\rm km\,s^{-1}}$ (including negative $v_{r}$) is generally non-outflowing gas at the cooler end of the temperature distribution. 
At higher radial velocities, the thermal emission tends to come preferentially  
from higher-temperature gas.   We show below that in the very high-velocity wind, much of the mass is in the form of warm gas, but the thermal emission shown here is primarily sensitive to the hotter material. 

We caution that the dominant cooling emission for gas with $T_{\rm gas}\sim10^{6}-10^{7}\,$K probably comes from metal lines, not thermal bremsstrahlung. Our treatment of enrichment in the current models is relatively simple and just traces the total metallicity (improved models will be presented in future work in preparation); therefore it is difficult to make detailed predictions for the line emission. However, we can use the compiled tables presented in \citet{wiersma:2009.coolingtables}, who use a suite of CLOUDY calculations to determine the total metal-cooling luminosity as a function of $n_{e}$, $n_{i}$, $Z$ (assuming solar abundance ratios), $T_{\rm gas}$ and the photo-ionizing UV background (for simplicity the $z=0$ background is used), to calculate the metal cooling luminosity. We plot the morphology of this emission as well in Figure~\ref{fig:wind.phase.morph}. It is broadly similar to the thermal bremsstrahlung emission, as expected; however, the metal-line emission is significantly more clumpy (and in the Sbc case more concentrated in vertical filamentary structures as well) -- closer to what is actually observed in systems like M82. This is because the metals are inhomogeneous. We caution, however, that metal diffusion and other small-scale mixing processes are not followed here, so the degree of inhomogeneity may be over-estimated. 

\subsection{Velocity and Column Density Distributions}
\label{sec:winds:phases.vel}

Figure~\ref{fig:wind.phases} (middle panel) also plots the radial velocity distribution 
of gas in the simulations, specifically the mass per unit radial velocity $dm/d{v_{r}}$.
There is a concentration of gas at small $|v_{r}|$ which is 
the rotationally-supported disk (this includes almost all the star-forming gas). 
This ``spike'' is symmetric about $v_{r}=0$; the winds appear as the 
large excess of material at $v_{r}\gg0$, extending 
to $v_{r} > 1000\,{\rm km\,s^{-1}}$ even in the SMC-like case. 

In each galaxy model, the winds span a wide range in outflow velocity -- there is no 
single ``wind velocity'' (contrary to what is assumed in many models).
The distribution of mass in $dm/d{v_{r}}$ has a broad 
plateau up to some ``turnover'' velocity, above which it falls off more rapidly; 
we can approximate it with 
\be
\label{eqn:windmass.vs.vr}
\frac{dm_{\rm wind}}{d{v_{r}}} \propto \frac{1}{1+(v_{r}/v_{0})^{\alpha}}
\ee
with $\alpha\sim3-7$ and $v_{0}\sim800,\,500,\,200,$ and $\,200\,{\rm km\,s^{-1}}$ 
in the HiZ, Sbc, MW, and SMC cases, respectively. For comparison, the maximum circular velocities of these systems are $V_{\rm max}\approx\,230,\,86,\,190$ and $46 \ {\rm km\,s^{-1}}$. In the 
starburst systems where radiation pressure is critical for much of the wind (HiZ/Sbc), 
the turnover velocity is a few times $V_{\rm max}$ (comparable to the 
escape velocity) -- this is expected for radiatively accelerated winds. 
In contrast, the maximum velocities in the lower-SFR systems appear to asymptote to a maximum 
at a few hundred ${\rm km\,s^{-1}}$ -- such a roughly constant velocity is, in turn,  the expectation for thermally heated galactic outflows.

\begin{figure}
    \centering
    \plotone{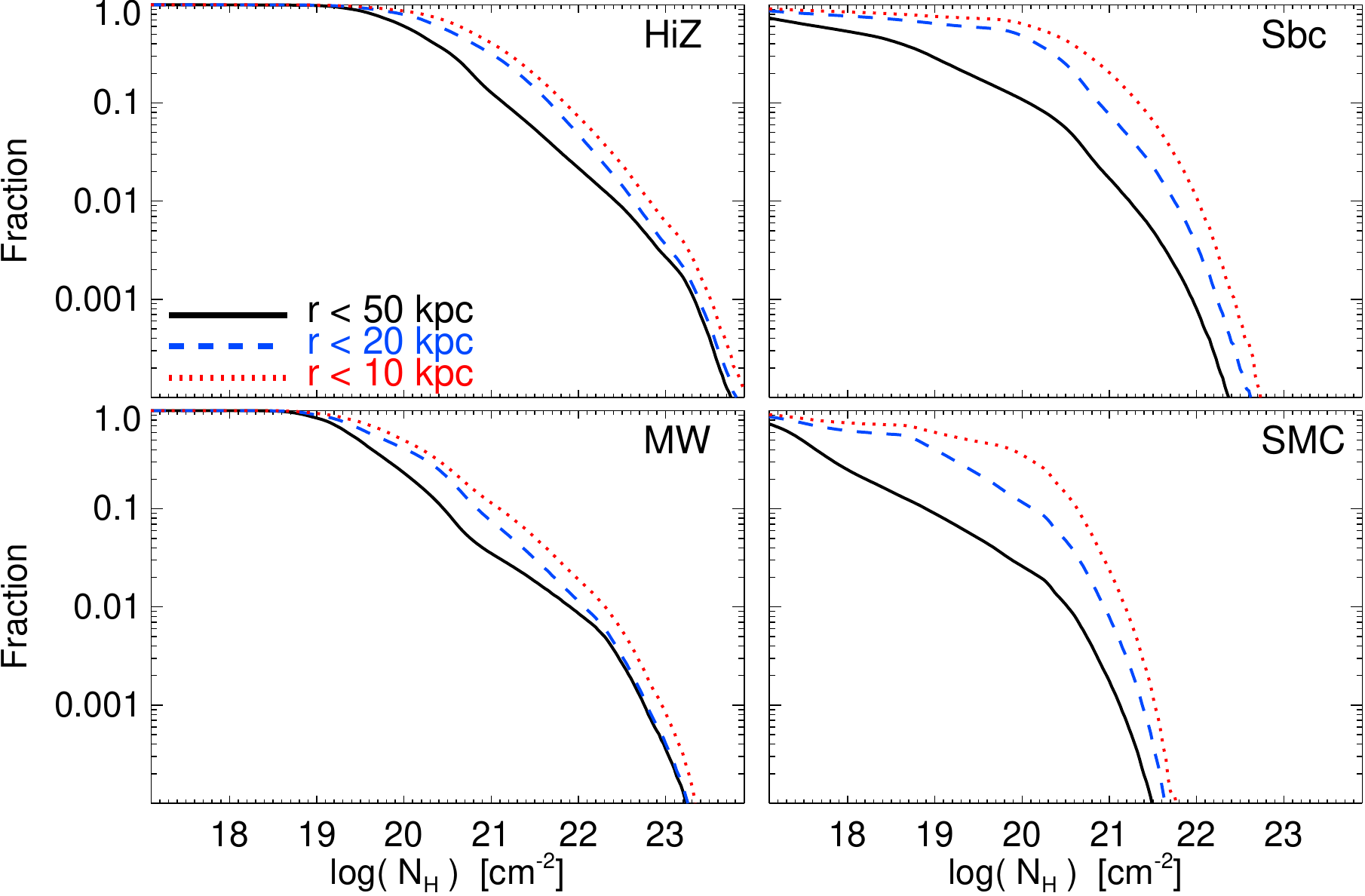}
    \caption{Cumulative distribution of gas column density (fraction $>N_{H}$) along random sightlines within different 
    radii around each galaxy in our simulations with all feedback mechanisms.  The results are  averaged over time and projected disk angle, for random sightlines 
    within a projected circular aperture of physical radius $r$ around the disk center (labeled).
    The few percent of sightlines at large columns ($\gtrsim10^{22}\,{\rm cm^{-2}}$) 
    go through the disk itself; the remainder are dominated by 
    wind material. Wind material with $N_{H}\gtrsim10^{19}\,{\rm cm^{-2}}$ 
    has a large covering factor within a few tens of kpc, 
    and there is a broad range of columns $\sim10^{18}-10^{21}\,{\rm cm^{-2}}$.
    \label{fig:wind.columns}}
\end{figure}

Observationally, the column density distribution of the gas is probed directly by absorption line measurements (e.g., Na and Mg lines).   Figure \ref{fig:wind.columns} quantifies this distribution in our simulations.  To calculate this, we determined the average column density 
distribution calculated over $\sim100$ different projections of the galaxy, 
each time calculating the integrated line-of-sight column along $\sim10^{4}$ sightlines through an aperture of fixed physical radius about the galaxy center (uniformly sampling the aperture area). Note this calculation is simply the total hydrogen column density -- we do not include the ionization corrections that are critical to computing realistic line strengths. The very large columns in Figure \ref{fig:wind.columns}
come from the few percent of sightlines which go through the disks, but the 
bulk of the column density distribution is dominated by the winds. Figure \ref{fig:wind.columns} shows that there is a broad range of columns with a roughly power-law behavior in $N_{H}$ at low values (expected for an expanding, constant $\dot{M}$ wind); 
the characteristic columns range from $\sim10^{18}-10^{20}\,{\rm cm^{-2}}$ around dwarf galaxies 
to $\sim10^{20}-10^{22}\,{\rm cm^{-2}}$ around massive systems.

\subsection{Cold, Warm, and Hot Outflows}
\label{sec:winds:phases.coldwarmhot}

The mass distribution as a function of radial velocity in Figure~\ref{fig:wind.phases} divides the gas into three traditional phases with a simple 
temperature cut: ``cold atomic+molecular'' gas ($T<2000\,K$), ``warm ionized'' gas
($2000<T<4\times10^{5}\,K$) and ``hot'' gas ($T>4\times10^{5}\,K$). 
The wind is a mix of warm gas (containing most of the mass) and volume-filling hot 
gas, with a few percent contribution from cold gas at all velocities. 
This is shown directly in the lower panel of Figure~\ref{fig:wind.phases}, 
where we consider the density distribution of wind material (defined as all unbound 
gas, see below), divided into these same phases. The 
wind includes material with a broad range of densities $n\sim 10^{-6}-10^{2}\,{\rm cm^{-3}}$, 
with high-density material at $n\gtrsim 1\,{\rm cm^{-3}}$ primarily ``cold,'' 
intermediate-density material ($n\sim 0.01-1\,{\rm cm^{-3}}$) ``warm,'' and 
low-density material ($n\lesssim 0.01\,{\rm cm^{-3}}$) ``hot.''

The instantaneous phase structure in the wind is not the same, however, as the phase 
structure of the material at the time it was initially accelerated into the wind. 
Its multi-phase origin is evident in the fact that the 
warm gas is highly inhomogenous; most of the mass is at a density $\sim20-100$ times higher than would be expected if it were smoothly distributed over a spherical shell (the formal clumping factor is 
$\langle \rho_{\rm }^{2}\rangle/\langle \rho_{\rm }\rangle^{2}\approx100$ in each case).
Some of the warm gas -- especially in the larger and more diffuse loops and shells, is material from the diffuse ``warm-phase'' of the disk disk (which has both sizeable mass and volume filling factors) directly accelerated out of the disk. But the large tail of very low-density ($n\lesssim10^{-4}\,{\rm cm^{-3}}$) volume-filling warm gas is 
material initially launched as ``hot'' gas, which cools adiabatically as it expands.
And much of the dense clumpy/filamentary structure is material initially launched as ``cold'' gas, 
which rapidly becomes photo-ionized as it expands and its average density falls.

There are several important numerical caveats to the micro-scale phase structure in the wind shown here.  In particular, acceleration of cold gas directly by hot gas is difficult to treat accurately in SPH codes, and comparison with adaptive mesh schemes suggests that we are likely to under-estimate the true amount of cold material in the wind
\citep{springel:arepo,keres:2011.aas.arepo}. Similarly, we do not capture the thermal evaporation of cold gas to form hot gas or the condensation of hot gas to form cool gas.  And the adiabatic expansion of the gas to extremely low densities is possible because our simulations do not include a realistic IGM into which the gas would expand.
As a result, the phase distributions shown in Figure~\ref{fig:wind.phases} are unlikely to be quantitatively accurate.

\begin{figure*}
    \centering
    \plotside{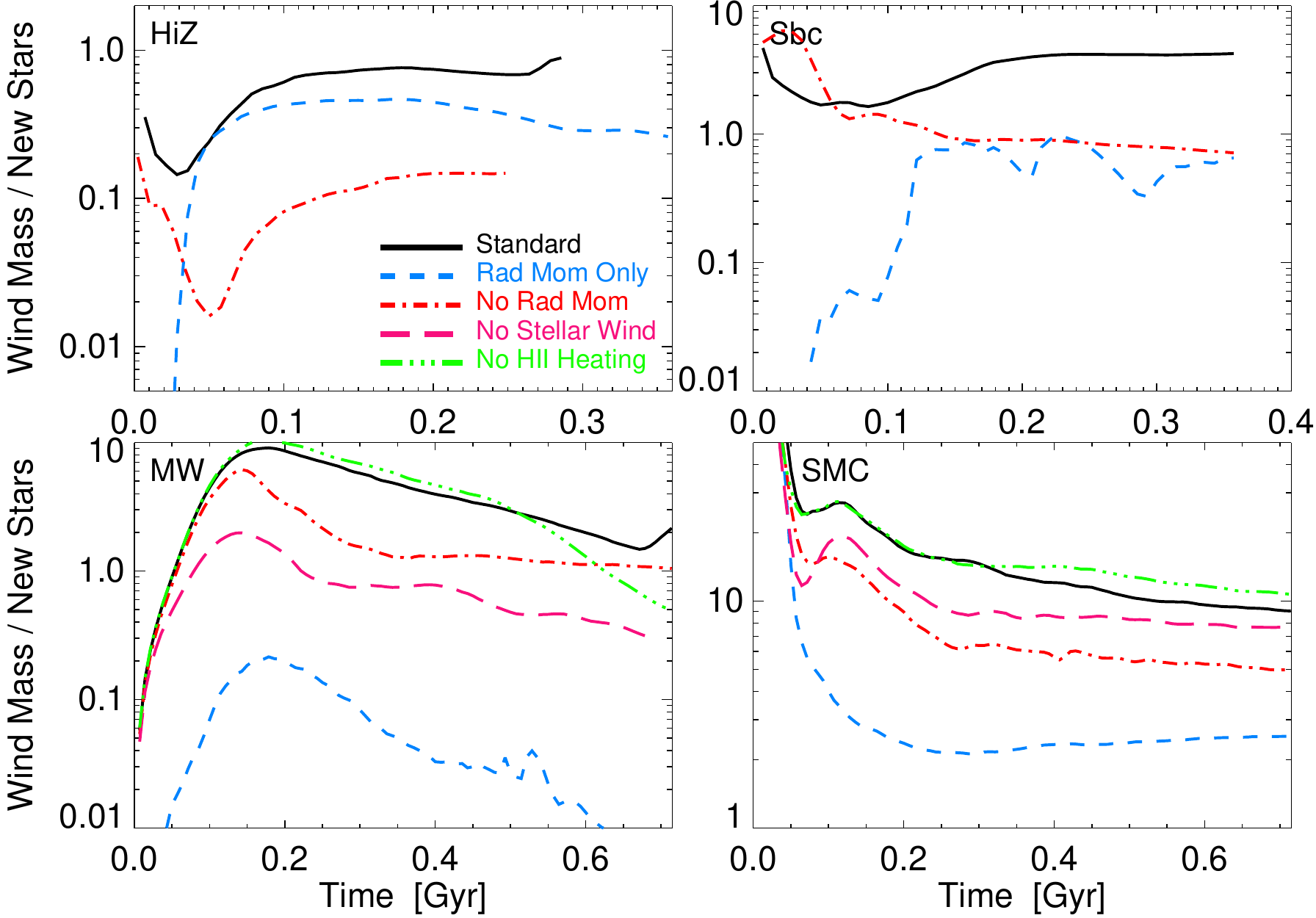}
    \caption{Galactic wind mass-loading efficiency ($\equiv M_{\rm wind}/M_{\rm new}$; 
    where $M_{\rm wind} = \int\,\dot{M}_{\rm wind}$ and $M_{\rm new} = \int\,\dot{M}_{\ast}$) 
    for each galaxy model. We compare our standard simulations with all feedback mechanisms
    to simulations with different feedback mechanisms enabled in turn. 
    Models with no feedback 
    produce no measurable wind mass.   Since SNe dominate the production of wind material by hot gas (relative to HII heating and stellar winds), the ``Rad Mom Only'' models shown here are very similar to ``no SNe'' models.     {\em Top Left:} HiZ model. Net outflow rate is $\sim \dot{M}_{\ast}$.
    Most of the wind comes from 
    ``cold'' acceleration by radiation pressure, and is present even with 
    no SNe/stellar wind/HII heating terms (though these do contribute a non-negligible fraction 
    from ``venting,'' seen in the no radiation pressure case). 
        {\em Top Right:} Sbc model, with $\dot{M}_{\rm wind}\sim 3-5 \, \dot{M}_{\ast}$. 
    Heating mechanisms contribute comparably to radiation pressure 
    in driving the wind. Note that the ``standard'' model wind mass is significantly larger than the 
    sum of the wind mass with just heating and just radiation pressure enabled separately -- there is a strong non-linear coupling between the different feedback mechanisms that enhances the strength of the wind.
    {\em Bottom Left:} MW-like model, with $\dot{M}_{\rm wind}\sim 1-10\, \dot{M}_{\ast}$. 
    Here the hot gas produced by SNe dominates the wind driving. 
        {\em Bottom Right:} SMC-like model, with $\dot{M}_{\rm wind}\sim 10-20 \, \dot{M}_{\ast}$.
    Hot gas is again critical, with SNe even more dominant.
    \label{fig:wind.vs.fb}}
\end{figure*}


\section{Mass-Loading of Galactic Winds}
\label{sec:winds}

\subsection{Effects of Each Feedback Mechanism in Turn}
\label{sec:winds:vsfb}

From the broader perspective of galaxy formation, the most important integral property of galaxy winds is probably the total wind mass outflow rate.  
This is typically quantified in terms of its ratio to the star formation rate 
$\dot{M}_{\rm wind}/\dot{M}_{\ast}$, i.e.\ the wind mass per unit mass of stars formed.
Figure~\ref{fig:wind.vs.fb} shows the wind masses of 
our models, in a series of otherwise identical simulations with different feedback 
mechanisms enabled or disabled in turn.   This quantifies the importance of a given feedback process in driving the wind in our simulations.   We compare the total  mass in the wind $M_{\rm wind}$ as a function of time, relative to the integrated mass in stars formed since the start of the simulation $M_{\rm new}$ -- the ratio of these quantities $M_{\rm wind}/M_{\rm new}$ defines the average ``global'' wind efficiency $\langle \dot{M}_{\rm wind} \rangle/\langle \dot{M}_{\ast}\rangle$.  
If we evaluate the instantaneous efficiency or average in narrow time bins, we obtain similar results but with increasing scatter.

 In these calculations, we define the ``wind'' as material 
 with a positive Bernoulli parameter $b\equiv (v^{2} + 3\,c_{s}^{2} - v_{\rm esc}^{2})/2$, i.e., material that would escape in the absence of additional forces or cooling.\footnote{We 
 have also calculated the wind mass using a more observationally accessible definition: 
 material that has reached at least $>500\,$pc above the disk, with absolute outflow 
 velocity $>100\,{\rm km\,s^{-1}}$.   This gas may or may not all be formally unbound, and the exact ``cuts'' in $z$ and $v$ are arbitrary, but we find that this definition of wind mass is generally within $\sim20\%$ of the wind mass defined with the Bernoulli parameter.  Moreover, the results are essentially unchanged for moderate (factor $\sim2$) 
changes in the threshold values of $z$ and $v$.}

Figure~\ref{fig:wind.vs.fb} shows that for the HiZ galaxy model, the photon momentum plays a critical role in driving much of the wind mass.   By contrast, removing the sources of hot gas (SNe, stellar winds, and HII heating) has little effect on the total wind mass.   With all feedback mechanisms operating, a multi-phase wind is blown with an outflow rate between $\sim0.2-1$ times the SFR. 
Without local and long-range acceleration by photon momentum, however, the wind efficiency drops by a factor of $\sim 5$, both because the absolute wind mass drops and because the star formation rate increases signficantly.  The long-range acceleration due to the ambient radiation field is especially critical in driving the wind: we show in \papertwo\ 
that a moderate increase in the UV escape fraction from the galaxy can increase the wind mass by factors of $\sim2-5$.  By contrast, removing the local IR radiation pressure support within star forming regions has a much weaker effect on the wind (though we have shown in 
\paperone\ that it is critical for the support of GMC complexes themselves). The winds in the HiZ model therefore seem not to be directly launched from GMCs, but rather are the result of a two step process \citep{murray:2011.cluster.wind.launching}. First, UV and IR radiation pressure lofts up gas parcels. Second, these parcels are then continuously  accelerated by photons emerging from the entire disk, rather than only those from their natal GMC.

For the MW model, our fiducial calculations produce a wind with an outflow rate a couple times the SFR.\footnote{The high wind mass-loading at early times ($\dot M_{\rm wind} \sim 10 \dot M_\ast$) is partially an artifact of the initial conditions. This occurs in part because it takes
$\sim2\times10^{8}\,{\rm yr}$ for the SFR to reach equilibrium, and partly because 
the pre-simulation bulge+disk stars are initialized with a broad age range, so a small fraction are sufficiently young 
to have a large feedback effect before the SFR rises much (and because the galaxy is so gas-poor, this has as a proportionally large effect).}
  The relative role of the hot wind phase is much more prominent than in the HiZ case. If we remove all of the thermal feedback (SNe, stellar winds, and HII regions), 
there is essentially no wind!   And of the heating mechanisms, SNe have the largest effect on the wind properties.    Absent thermal heating, the SFR is still regulated at nearly the same 
level by the radiation pressure feedback (as in \paperone), but material is not being driven 
into a super-wind.  Radiation pressure thus plays an important role in self-regulating star formation in the disk, but is not, unlike in the HiZ models, strong enough to directly accelerate much gas far out of the disk.  Rather, radiation pressure  lofts material out of GMCs, after which the gas is accelerated primarily by the hot gas pressure 
into a galaxy-scale outflow. This contrasts with the HiZ case, where the acceleration above the disk results from radiation pressure rather than from hot gas ram pressure. 



The Sbc case, as before, lies somewhere in between the HiZ and MW cases. 
The net efficiency is a few times the SFR, with 
radiation pressure and hot gas contributing comparably to the outflow mass. 
Note also that Figure~\ref{fig:wind.vs.fb} clearly shows 
that the total wind mass in the case with all feedback enabled (``standard'') is 
significantly larger than the sum of the wind mass with hot gas alone or radiation pressure 
alone. The full wind efficiency stems from the non-linear interaction between 
different feedback mechanisms. 

For our SMC-like dwarf galaxy model, the simulation with all feedback mechanisms included 
drives an outflow of $\sim10-20$ times the SFR (Fig.~\ref{fig:wind.vs.fb}).   Just as in the MW model, SNe are the most important driver of the galactic outflow.  In particular if we remove all sources of hot gas (SNe and stellar winds), we again see essentially no significant super-wind. 
By contrast, without any feedback from radiation pressure, the wind mass is reduced 
by a small amount ($\sim30-40\%$) and the average SFR increases by a similar factor, so 
the net outflow efficiency is only a factor $\sim1.7$ lower.

\begin{figure*}
    \centering
    \plotside{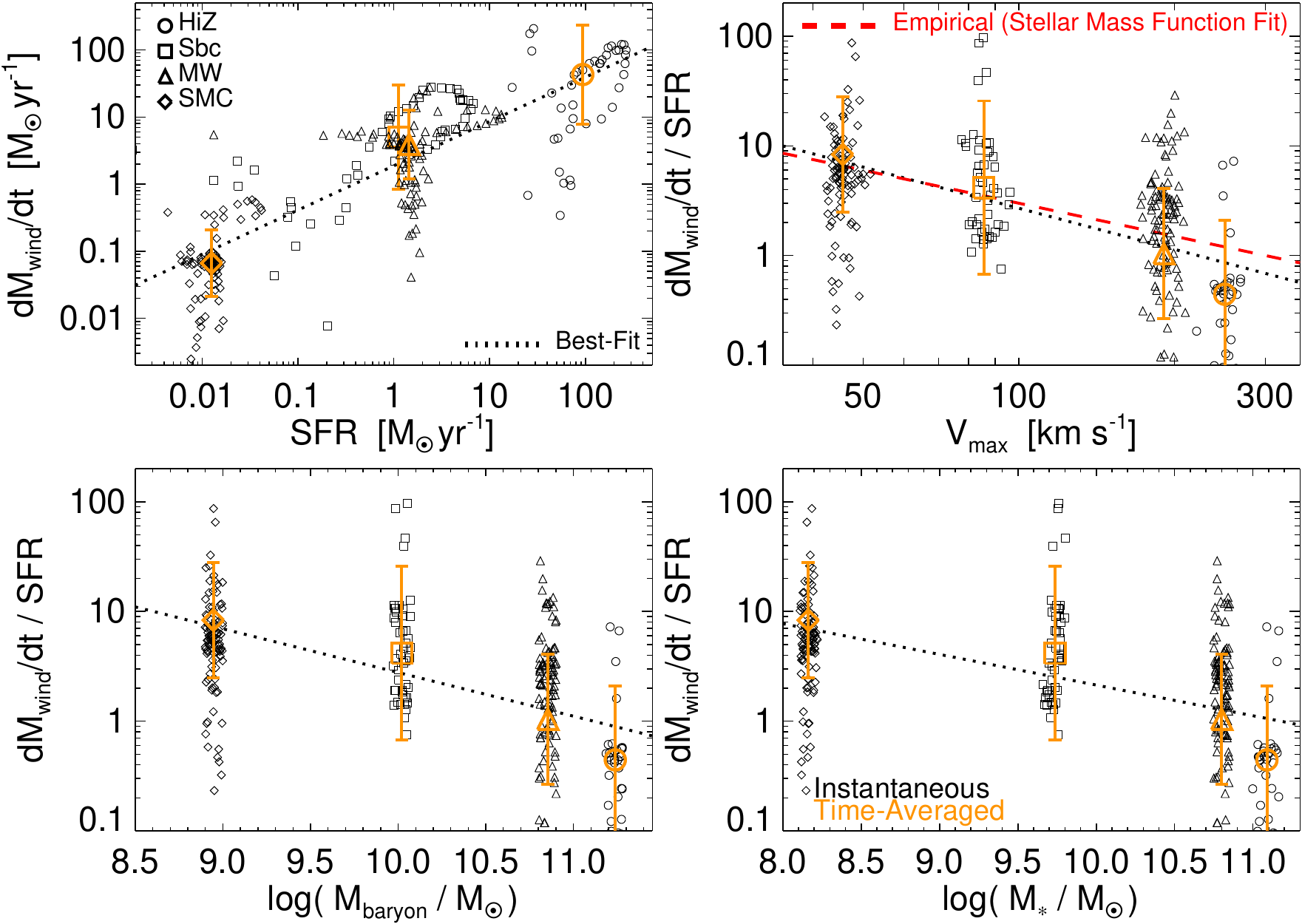}
    \caption{Summary of the efficiency of stellar winds, defined as the wind mass loss rate
    divided by the SFR, $\dot{M}_{\rm wind}/\dot{M}_{\ast}$. 
    Each of our galaxy models is shown, for simulations with all feedback mechanisms included.  The instantaneous efficiencies (the derivative of 
    Fig.~\ref{fig:wind.vs.fb}) are shown at a number of times (black); the time-averaged 
    value and $\pm1\,\sigma$ scatter over the duration of the simulation is also shown (orange 
    with error bars).
    The best-fit power-law relation is also plotted in each panel (dotted line).
    {\em Top Left:} Outflow rate versus SFR. The best-fit is sub-linear: 
    $\dot{M}_{\rm wind}\propto \dot{M}_{\ast}^{0.66\pm0.08}$ (eq.~\ref{eqn:mdot.wind.vs.sfr}). 
    {\em Top Right:} Efficiency versus the maximum circular velocity of  the galaxy rotation curve. 
  Our best-fit is $\dot{M}_{\rm wind}/\dot{M}_{\ast}\propto V_{\rm max}^{1.2\pm0.2}$
    (eq.~\ref{eqn:mdot.wind.vs.vmax}).   For comparison, we also show the scaling from cosmological simulations that are adjusted to match high-redshift IGM enrichment and approximately match the observed galaxy mass function \citep{oppenheimer:outflow.enrichment,oppenheimer:recycled.wind.accretion}:$^{\ref{foot:oppenheimernote}}$
    $\langle\dot{M}_{\rm wind}/\dot{M}_{\ast}\rangle = (V_{\rm max}/{300\,{\rm km\,s^{-1}}})^{-1}$(red dashed).  This is consistent within $1\,\sigma$ with our numerical results. 
    {\em Bottom Left:} Wind efficiency versus galaxy baryonic mass. 
    {\em Bottom Right:} Wind efficiency versus galaxy stellar mass. 
    \label{fig:wind.summary}}
\end{figure*}

\begin{figure}
    \centering
    \plotone{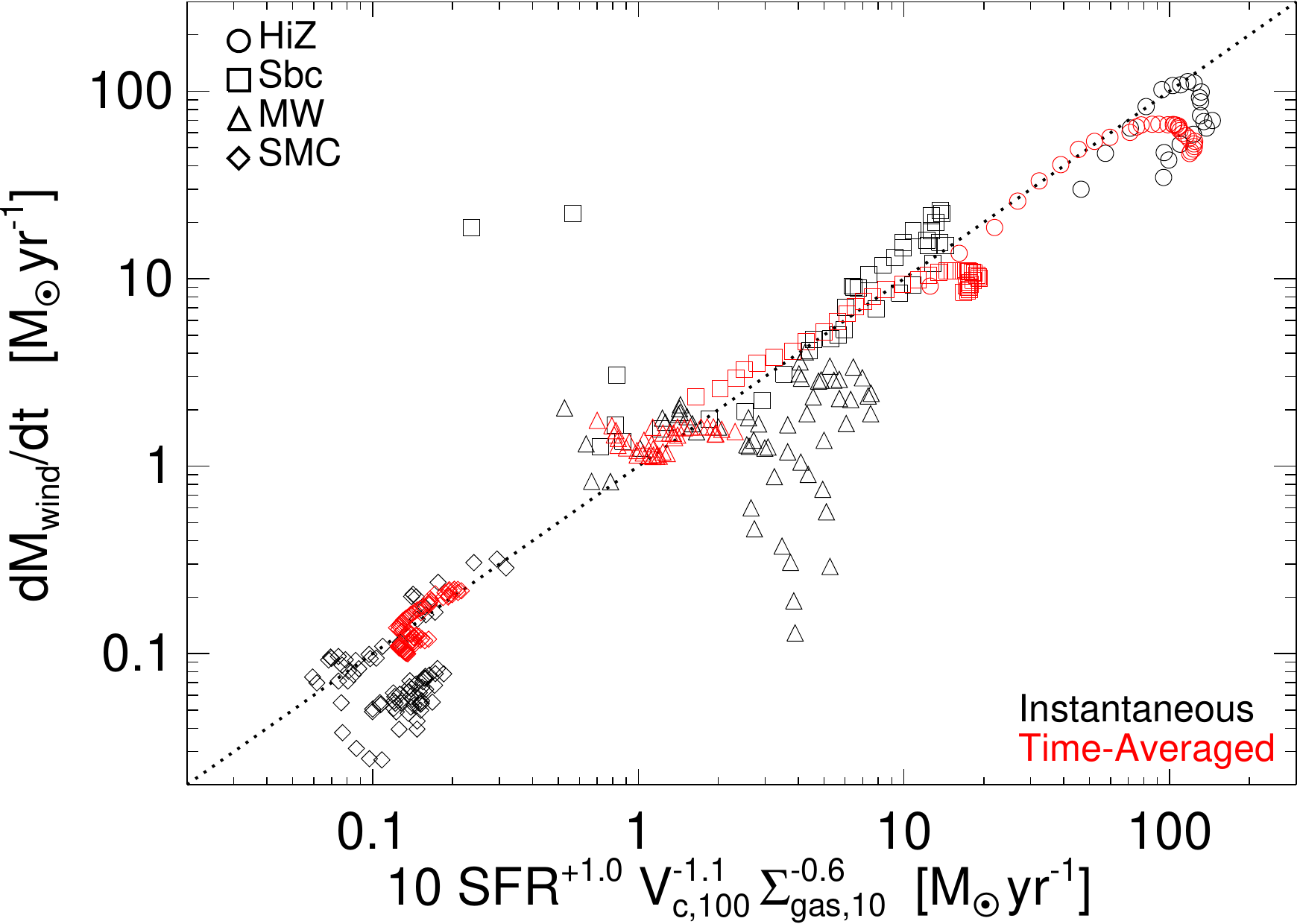}
    \caption{Best-fit scaling of the wind mass rate as a function of galaxy properties in our numerical simulations (eq.~\ref{eqn:bestfit.eqn}):  roughly $\dot M_{\rm wind}/\dot M_\ast \propto V_c^{-1} \, \Sigma_{\rm gas}^{-1/2}$ (where the quantities are evaluated at each wind launching radius $R$ 
    in the galaxy). The dotted line shows the identical relation. Black points 
    show the instantaneous efficiencies (as Fig.~\ref{fig:wind.summary}), red points 
    the running time-average (as Fig.~\ref{fig:wind.vs.fb}, with
    quantities evaluated at the half-SFR radius at each time).    
    The order-of-magnitude scatter in Fig.~\ref{fig:wind.summary} is reduced to factor $\sim2$.  The scaling $\dot M_{\rm wind}/\dot M_\ast \propto \Sigma_{\rm gas}^{-1/2}$ reflects the fact that in denser systems, SNe cool more efficiently and more of the optical/UV radiation is absorbed, reducing the efficiency of wind driving. 
    \label{fig:wind.bestfit}}
\end{figure}

\subsection{Wind Efficiency versus Galaxy Mass}
\label{sec:winds:vsmass}

Figure~\ref{fig:wind.summary} summarizes the efficiency of the winds in our 
standard simulations as a function of global galaxy properties. 
We plot the wind efficiency at several different times 
(evaluated every $10^{7}\,$yr), for each standard disk model, 
as a function of several parameters: the instantaneous SFR, 
the maximum circular velocity of the disk $V_{\rm max}$, and the 
stellar and baryonic galaxy masses. 
The order-of-magnitude scatter in wind efficiencies even at fixed galaxy properties 
is clear, but there are also clear trends in the average efficiencies with each 
galaxy property. 

The wind mass loading is strongly correlated with the galaxy SFR, but 
with a sub-linear power. Fitting a single power-law in Figure~\ref{fig:wind.summary} 
gives 
\be
\label{eqn:mdot.wind.vs.sfr}
\dot{M}_{\rm wind} \sim 3\,\msun\,{\rm yr^{-1}}\,{\Bigl(} 
\frac{\dot{M}_{\ast}}{\msun\,{\rm yr^{-1}}} {\Bigr)}^{0.7}
\ee
The sub-linear behavior is particularly interesting, as is implies 
a lower efficiency at higher absolute SFR: 
$\dot{M}_{\rm wind}/\dot{M}_{\ast}\propto \dot{M}_{\ast}^{-0.3}$. 
Since observations find a SFR-stellar mass relation 
$\dot{M}_{\ast}\propto M_{\ast}^{0.7}$ in star-forming systems \citep{noeske:2007.sfh.part1}, this 
is equivalent to a lower efficiency at higher masses 
($\dot{M}_{\rm wind}/\dot{M}_{\ast}\propto M_{\ast}^{-0.23}$).

This is evident when we directly plot the efficiency versus  baryonic mass, stellar mass, or maximum circular velocity.
The strongest correlation appears to be between efficiency and 
$V_{\rm max}$. 
This has the form 
\be
\label{eqn:mdot.wind.vs.vmax}
{\Bigl \langle} \frac{\dot{M}_{\rm wind}}{\dot{M}_{\ast}} {\Bigr \rangle}
\sim 3[\pm1]
{\Bigl(} 
\frac{V_{\rm max}}{100\,{\rm km\,s^{-1}}}
{\Bigr)}^{-(0.9-1.7)}\ .
\ee
This is in reasonable agreement with the expectation from simple dimensional 
arguments based for a momentum-driven wind 
with a characteristic velocity of $V_{\rm max}$ \citep{murray:momentum.winds}. 
If the momentum flux in the outflow scales with the luminosity  of
massive stars ($\propto L \propto \dot{M}_{\ast}$), 
then the outward momentum $\dot{M}_{\rm wind}\,V_{\rm esc}\propto \dot{M}_{\ast}$, 
so $\dot{M}_{\rm wind}/\dot{M}_{\ast} \propto V_{\rm max}^{-1}$.   Note that this scaling can also apply if ram pressure due to hot gas is the primary mechanism driving mass (i.e., warm/cold gas) out of the galaxy.\footnote{We should note that, although it is not obvious because of the large scatter in instantaneous mass-loading rates, 
the best-fit trend in Figure~\ref{fig:wind.summary} is robust to the exclusion of any individual simulation.}
There is a small, but real deviation of the HiZ model from the extrapolation of this scaling, which is a consequence of an additional residual dependence on $\Sigma_{\rm gas}$ that we discuss below.

The scaling of $\dot M_{\rm wind}/\dot M_{\ast}$ with $V_{\rm max}$ in Figure~\ref{fig:wind.summary} and equation \ref{eqn:mdot.wind.vs.vmax}
is in plausible agreement with the scaling suggested in cosmological 
simulations in order to produce a good match between 
the predicted and observed galaxy stellar (and baryonic) mass functions as well as high-redshift IGM enrichment patterns. 
\citet{oppenheimer:outflow.enrichment} and \citet{oppenheimer:recycled.wind.accretion} 
\citep[see also][]{dutton:2009.wind.scalings.needed} 
considered a series of prescriptions for how galactic winds scale with galaxy properties, and ultimately 
found agreement with observations for a scaling 
$\dot{M}_{\rm wind}/\dot{M}_{\ast} = 3\,(V_{\rm max}/100\,{\rm km\,s^{-1}})^{-1}$, 
which lies in the best-fit range of our simulations (as shown by the red line in Fig.~\ref{fig:wind.summary}).\footnote{\label{foot:oppenheimernote}\citet{oppenheimer:outflow.enrichment,oppenheimer:recycled.wind.accretion} actually do not resolve $V_{\rm max}$ and instead adopt a proxy for the velocity dispersion $\sigma$ based on a galaxy mass estimated from a friends-of-friends object finder. Without a matching cosmological simulation, it is difficult to exactly convert between this and the quantities measured here, so there is some ambiguity in whether the normalization estimated in the quoted values agrees exactly with that implemented. The scaling with $V_{\rm max}^{-1}$, however, should be more robust.}
We stress that we have not tuned or adjusted any parameters in our simulations to produce either 
the normalization or slope of this relation.

Qualitatively similar behavior appears when we consider the wind
efficiency versus baryonic or stellar mass, albeit with a slightly 
less significant correlation. 
\begin{align}
\label{eqn:mdot.wind.vs.mbaryon}
{\Bigl \langle} \frac{\dot{M}_{\rm wind}}{\dot{M}_{\ast}} {\Bigr \rangle}
& \sim 3.5[\pm1]
{\Bigl(} 
\frac{M_{\rm baryon}}{10^{10}\,\msun}
{\Bigr)}^{-(0.3-0.65)} \\ 
\label{eqn:mdot.wind.vs.mstellar}
{\Bigl \langle} \frac{\dot{M}_{\rm wind}}{\dot{M}_{\ast}} {\Bigr \rangle}
& \sim 2[\pm 0.5]
{\Bigl(} 
\frac{M_{\ast}}{10^{10}\,\msun}
{\Bigr)}^{-(0.25-0.5)} \ .
\end{align}
Given the observed
(stellar and baryonic) Tully-Fisher relation ${M}_{\ast}\propto V_{\rm max}^{4}$; 
these correlations are fully consistent with the $V_{\rm max}$ correlation; 
likewise given the $\dot{M}_{\ast}-M_{\ast}$ relation discussed above, 
these are consistent with the wind mass-SFR relation. 
This consistency is expected, given that the simulated galaxies 
lie on both the observed Tully-Fisher relations (by design) and 
SFR-stellar mass relation. 

\subsection{Predicting Wind Efficiencies: Towards a More Accurate Effective Model}
\label{sec:winds:prediction}

In this section we use the results of our simulations to assess if there is  
a more accurate predictor of wind properties than equations \ref{eqn:mdot.wind.vs.sfr}-\ref{eqn:mdot.wind.vs.mstellar} -- such a predictor would be extremely useful for application in cosmological simulations and/or semi-analytic models.

First, we consider the simulations alone, including all of our galaxy models 
and case studies where we have varied other parameters such as the 
galaxy structural properties and how the SFR is tied to molecular gas.
We isolate a series of properties that are measurable in such models 
(each as a function of radius): SFR, circular velocity, escape velocity, 
total/bulge/disk/gas/halo masses and surface densities, scale-heights, 
and the radii themselves (most other properties can be constructed from 
combinations of these). We then compare to the corresponding 
instantaneous wind mass loading $\dot{M}_{\rm wind}(R,t)$. 
Although all the parameters are correlated, if we marginalize over the 
other variables we find three parameters that dominate the trends 
in $\dot{M}_{\rm wind}$ with radius, time, and galaxy model: 
the SFR $\dot{M}_{\ast}$, circular velocity\footnote{Because 
we are considering the winds at 
different galactic radii $\dot{M}_{\rm wind}(R,t)$, we use $V_{c}(R)$ instead of  
$V_{\rm max}\equiv {\rm MAX}(V_{c}[R])$.} $V_{c}$, and gas 
surface density $\Sigma_{\rm gas}$.
Accounting for these, 
we do not find $>3\,\sigma$ dependence on any other variables we consider.
A maximum-likelihood fit to these three variables gives a correlation of the form 
\be
\dot{M}_{\rm wind}\propto 
\dot{M}_{\ast}^{\alpha}\,V_{c}^{\beta}\,\Sigma_{\rm gas}^{\gamma}
\label{eqn:bestfit}
\ee
with $\alpha\approx 1.0\pm0.15$, 
$\beta\approx-1.1\pm 0.25$, 
and $\gamma\approx -0.5\pm0.15$. 
From this, the correlation with SFR is most significant 
($>5\,\sigma$), and at otherwise fixed properties scales linearly ($\dot{M}_{\rm wind}\propto \dot{M}_{\ast}$); 
the sub-linear behavior in Fig.~\ref{fig:wind.summary} owes mostly to the next most-significant correlation, 
with $V_{c}(R,\,t)$ ($\sim4\,\sigma$), which is consistent with the expected $\propto V_{c}^{-1}$. 
Third most important (but still $>3\,\sigma$) is a residual correlation with surface density 
which dominates many of the variations in $\dot{M}_{\rm wind}/\dot{M}_{\ast}$ within 
a given model as a function of radius and time.   If we include the normalization in the fit, we find the following result for the mass loading \be
\label{eqn:bestfit.eqn}
{\Bigl \langle}
\frac{\dot{M}_{\rm wind}}{\dot{M}_{\ast}}
{\Bigr \rangle}_{R}
\approx 
10\, \eta_{1}\,
{\Bigl (} \frac{V_{\rm c}(R)}{100\,{\rm km\,s^{-1}}} {\Bigr)}^{-(1+\eta_{2})}\,
{\Bigl (} \frac{\Sigma_{\rm gas}(R)}{10\,{\rm \msun\,{\rm pc}^{-2}}} {\Bigr)}^{-(0.5+\eta_{3})}\,
\ee
where $\eta_{1},\,\eta_{2},\,\eta_{3}$ incorporate the uncertainty 
from the fits ($\eta_{1}\sim0.7-1.5$, $\eta_{2}\sim\pm0.3$, $\eta_{3}\sim\pm0.15$).  A comparison between this fit and the simulation results is shown in Figure \ref{fig:wind.bestfit}.  

Note that the scalings with star formation rate and circular velocity in equation \ref{eqn:bestfit} are consistent with the simple momentum conservation argument discussed after equation \ref{eqn:mdot.wind.vs.vmax}.   The surface density scaling is, however, new.  It implies that the wind mass loading is lower for higher gas surface density systems, all other things being equal.  Physically, our interpretation of this is as follows.   First, consider low mass systems where SNe dominate the wind driving.   As shock-heated SN bubbles expand, the momentum of the swept-up material increases in time due to the P-dV work done by the hot gas.   This enhances the wind driving so long as cooling does not sap the energy of the SN bubbles.  For higher surface density galaxies, however, radiative cooling of SN remnants is more important, suppressing the SNe contribution to the wind.  This is qualitatively consistent with the scaling in equation \ref{eqn:bestfit.eqn}.

Consider now the opposite limit of massive, dense systems in which radiation pressure is the most important wind driving mechanism in our models, in particular the long-range continuous 
acceleration due to the diffuse radiation field.    The gas at large heights above the midplane is generally not optically thick in the IR, but still has $\tau>1$ in the UV (and possibly in the optical, if the wind is sufficiently dense).  Thus the incident radiation pressure force on this gas depends on the escape fraction in the UV/optical  $f_{\rm esc,\,uv-opt}$ times $L/c$, so that $\dot{M}_{\rm wind}
\propto f_{\rm esc,\,uv-opt}\,\dot{M}_{\ast}\,V_{c}^{-1}$). In 
the Appendix, we explicitly confirm that the wind mass-loading in the HiZ and Sbc models indeed scales  close to linearly with $f_{\rm esc,\,uv-opt}$. 
The escape fraction in turn declines for larger disk gas surface densities, because 
the disk is itself what is absorbing the UV flux.  This implies a wind mass-loading that decreases for higher gas surface densities, again qualitatively consistent with the scaling in equation \ref{eqn:bestfit.eqn}.

\section{Discussion}
\label{sec:discussion}

We have studied the origin of galaxy-scale outflows using a new
set of numerical methods for modeling stellar feedback 
in hydrodynamic galaxy simulations; the feedback processes include
radiation pressure on the scales of star clusters; shock-heating, momentum injection, and gas recycling via supernovae and stellar winds; photo-ionization in (overlapping) HII regions; and radiation pressure  produced by the diffuse interstellar radiation field.   Our calculations use the results of stellar population modeling to self-consistently include the time-dependence of these feedback processes.      We have  explored this physics in the context of isolated (non-cosmological) galaxy models that range from those motivated by massive $z \sim 2$ galaxies forming stars at $\sim 100-300 \, \msunyr$ to models of SMC-like dwarf galaxies. 

By incorporating multiple feedback processes, we find that numerical simulations produce galaxy-scale outflows with mass loading factors up to  $\dot{M}_{\rm wind}\gtrsim10\,\dot{M}_{\ast}$.   Moreover, the wind mass-loading is the largest in the lowest mass galaxies, scaling approximately as $\sim V_{c}^{-1}$  (although with an additional gas surface density dependence that we discuss below).   This is, to our knowledge, the first time that hydrodynamic simulations have demonstrated that stellar feedback processes can self-consistently generate galactic winds with {\em sufficiently large mass-loading and the appropriate galaxy mass-dependence} to match what has been invoked to explain the shape of the the galaxy stellar mass function.

A  general feature of our results is  that the interaction between multiple feedback processes, acting on different spatial and temporal scales, is important for driving realistic galactic winds.   In particular, some feedback processes prevent runaway collapse within GMCs, thus maintaining a diffuse phase of the ISM that is more susceptible to feedback.  In general a different set of processes then accelerates the wind material out of the star-forming disk.   

By turning off individual processes in our numerical simulations, we can  identify the mechanisms responsible for most of the wind mass-loading in a given galaxy model.
In dense gas, typical of high-redshift 
rapidly star-forming galaxies (our HiZ model) and 
local starburst galaxies (local LIRGs and dense nuclei such as Arp 220), 
cooling times even in the ``diffuse'' ISM are extremely short, and thus energy input into the ISM by SNe and stellar winds is rapidly radiated away.    This heating can neither  halt 
the runaway collapse of massive GMC complexes  or accelerate significant amounts of material into a wind (since most of the gas is in the dense medium).    Instead, radiation pressure from trapped IR photons in star-forming regions serves to dissociate those regions, 
and ``lofts up'' gas above the disk -- there, the gas  ``feels'' the coherent momentum flux 
from the optical/UV/IR photons escaping the other star clusters in the disk, and can be continuously accelerated out of the galaxy. Most of the wind mass in our models of these dense systems is in cold, dense blobs radiatively 
accelerated out of the galaxy; the small amount of hot gas is volume filling but has little mass, and is too tenuous to significantly accelerate the dense material.  Under these conditions, we find that 
the net wind efficiencies are typically moderate, $\dot{M}_{\rm wind}\sim0.1-2\,\dot{M}_{\ast}$.

In lower-gas density systems such as the MW and dwarf starbursts such as M82 (analogous to our Sbc model) ($\langle n \rangle \sim 1\,{\rm cm^{-3}}$), heating mechanisms and 
direct momentum injection have comparable effects. 
Injection of momentum by radiation pressure, O-star winds, and photo-heating
when stars clusters have ages $<10^{6}\,$yr begins to dissociate their host GMCs and punch holes that allow SNe remnants to expand and escape into the diffuse/hot medium. There, 
cooling times are low, and bubbles can grow, overlap, and do $P\,dV$ work on the entrained 
warm and cold gas, contributing significantly to the wind driving.
With all feedback mechanisms included, dwarf starbursts generate galactic winds 
with mass-outflow rates $\sim2-5$ times the SFR. 

In our dwarf galaxy model (motivated by the SMC), and in the outer extended disks of more massive systems, the gas densities are very low ($\langle n \rangle \lesssim 0.1\,{\rm cm^{-3}}$). 
The cooling time can be comparable to or longer than the dynamical time so that
heating mechanisms become very efficient; such systems also have low stellar 
surface densities and gas optical depths, so radiation pressure is less 
important. A combination of momentum from UV photons, stellar winds, and warm gas 
pressure from photo-ionized regions stirs up star-forming regions, allowing SNe explosions to 
expand rapidly and generate large, overlapping hot bubbles. 
This generates a characteristic dwarf irregular morphology with patchy star-forming regions 
and hot SNe bubbles \citep[see also][]{governato:2010.dwarf.gal.form}.
These hot bubbles push cold shells and loops into a super-wind 
as they overlap, generating a multi-phase (mostly hot/warm) wind with a large 
mass loss rate $\dot{M}_{\rm wind}\sim 10-20\,\dot{M}_{\ast}$. 
Without SNe, there is still some hot gas and a ``bubble-like'' morphology driven by high velocity stellar (O-star) winds, but SFRs are a factor $\gtrsim5$ larger and (absolute) mass outflow rates a factor of $\sim20$ lower.


We have used our simulations to determine  improved prescriptions for wind mass-lass rates (eq.~\ref{eqn:bestfit.eqn}) and velocity distributions (eq~\ref{eqn:windmass.vs.vr}) for use in cosmological simulations and semi-analytic models of galaxy evolution (for which the explicit models here cannot yet be implemented):  $\dot M_{\rm wind}/\dot M_{\rm \ast} \propto V_c^{-1} \Sigma_{\rm gas}^{-1/2}$.   This fit is similar to what has been adopted phenomenologically in the past, but incorporates an additional correction such that the mass-loading decreases with increasing gas surface density $\propto \Sigma_{\rm gas}^{-0.5}$ (at fixed circular velocity).  In the radiation-pressure dominated regime, this reflects the fact that a dense disk captures more of the primary stellar luminosity very close to stars, downgrading photons into the IR. These degraded photons can help drive turbulence in very dense regions (with $\tau_{\rm IR}\gg1$) but generally do not contribute to accelerating material above the disk. In the SNe-dominated regime, the surface density scaling reflects the fact that more of the SN energy is radiated away in denser systems. 

In addition to introducing variations between galaxies with the same $V_{c}$, the $\Sigma_{\rm gas}$ dependence will introduce some additional systematic scaling with mass and circular velocity. At fixed redshift, total disk surface density tends to increase with mass and $V_{c}$, but at intermediate masses ($M_{\ast}\gtrsim10^{10}\,\msun$) gas fractions also decline with mass such that $\Sigma_{\rm gas}$ (measured within the star-forming disk, which is the important scale for winds) is nearly independent of galaxy mass \citep[e.g.][]{mcgaugh:tf,hall:2011.disk.scalings}. At lower masses, however, where $M_{\rm gas}\gtrsim M_{\ast}$, $\Sigma_{\rm gas}$ scales with mass as $\sim V_{c}^{1-2}$, steepening the mass-loading by an additional power of $V_{c}$. This may resolve some of the problems noted in \citet{dave:2011.mf.vs.z.winds}, where a simpler single power-law scalings tends to over-produce the number of low-mass galaxies.

The scaling we have inferred for the mass loss rate is applicable to a range of galaxies but may break down in two different extremes, which we will explore in future work. In the lowest mass dwarf galaxies (with $M_{\ast}\lesssim10^{6}\,\msun$ and $Z\ll0.01\,Z_{\sun}$), the ionizing background alone may be sufficient to suppress most structure, metallicities may be sufficiently low as to prevent molecule formation except under extreme conditions, and a single SNe can clear out the gas supply; our  simulations do not apply to these circumstances. At the opposite extreme, in the most dense regions of nuclear starbursts and material surrounding AGN, even the ``diffuse'' medium can be optically thick in the IR and many of our assumptions about the efficiency of radiation pressure and the (in)ability of IR photons to drive winds are likely to be modified.

We also caution that care is needed when mapping the scaling here -- which assumes the disk structure is marginally resolved and so quantities like $V_{c}$ and $\Sigma$ can be evaluated at the effective radius of star formation -- to models with poor resolution. For example, in the \citet{oppenheimer:outflow.enrichment} cosmological simulations 
(and many analytic models), $R_{\rm disk}$ and $V_{c}(R_{\rm disk})$ are 
unresolved, so they are implicitly assumed to scale in a simple manner with total galaxy 
baryonic mass (and redshift). If 
we use the observed Tully-Fisher relation and gas fraction-stellar mass relations as noted above 
to convert our proposed scaling to a strict power of mass 
and redshift, the scalings are similar at intermediate masses. But it is in general non-trivial to 
link the quantities relevant to the outflows in our simulations (e.g., Fig.~\ref{fig:wind.bestfit}) to global parameters like halo/baryonic mass or halo circular velocity. 

The calculations presented in this paper can clearly be improved in a number of ways in future work.   Most notably our treatment of the radiation pressure produced by UV, optical, and IR photons is quite approximate.    A full calculation including the scattering and absorption of both the UV and IR photons  is technically formidable.   We have calibrated our treatment of the diffuse radiation field in galaxies using such radiative transfer calculations (\papertwo) but additional improvements to our methods would be valuable.   Our calculations also do not include cosmic-rays generated in SN shocks, which empirically contribute significantly to the pressure-support of MW-like spiral galaxies \citep{boulares90} and may also contribute significantly to driving outflows from such galaxies. And we reiterate that even at our highest resolution, there are important numerical caveats (and unresolved mixing processes) that may apply to the phase structure of the winds.

Finally, our calculations have focused on idealized, isolated galaxy models and do not include a realistic hot coronal gas component, galactic inflow, or an external intergalactic medium.   The interaction between galactic outflows, hot halo gas, and inflowing cold streams is an extremely interesting problem that should be studied in detail in future work.   Because we have not captured this interaction, the outflow rates we calculate are best interpreted as outflow rates from the star-forming disk, rather than from the virial radius of the host dark matter halo.


\acknowledgments 
We thank Todd Thompson and Romeel Dav{\'e}
for helpful discussions.  Support for PFH was provided by the Miller
Institute for Basic Research in Science, University of California
Berkeley.  EQ is supported in part by NASA grant NNG06GI68G and the
David and Lucile Packard Foundation. NM is supported in part by NSERC
and by the Canada Research Chairs program.
\\

\bibliography{/Users/phopkins/Documents/lars_galaxies/papers/ms}

\begin{thebibliography}{62}
\expandafter\ifx\csname natexlab\endcsname\relax\def\natexlab#1{#1}\fi

\bibitem[{{Aguirre} {et~al.}(2001){Aguirre}, {Hernquist}, {Schaye}, {Weinberg},
  {Katz}, \& {Gardner}}]{aguirre:2001.igm.metal.evol.sims}
{Aguirre}, A., {Hernquist}, L., {Schaye}, J., {Weinberg}, D.~H., {Katz}, N., \&
  {Gardner}, J. 2001, \apj, 560, 599

\bibitem[{{Boulares} \& {Cox}(1990)}]{boulares90}
{Boulares}, A., \& {Cox}, D.~P. 1990, \apj, 365, 544

\bibitem[{{Brook} {et~al.}(2011)}]{brook:2010.low.ang.mom.outflows}
{Brook}, C.~B., {et~al.} 2011, \mnras, 415, 1051

\bibitem[{{Chen} {et~al.}(2010){Chen}, {Tremonti}, {Heckman}, {Kauffmann},
  {Weiner}, {Brinchmann}, \& {Wang}}]{chen:2010.local.outflow.properties}
{Chen}, Y.-M., {Tremonti}, C.~A., {Heckman}, T.~M., {Kauffmann}, G., {Weiner},
  B.~J., {Brinchmann}, J., \& {Wang}, J. 2010, \aj, 140, 445

\bibitem[{{Coil} {et~al.}(2011){Coil}, {Weiner}, {Holz}, {Cooper}, {Yan}, \&
  {Aird}}]{coil:2011.postsb.winds}
{Coil}, A.~L., {Weiner}, B.~J., {Holz}, D.~E., {Cooper}, M.~C., {Yan}, R., \&
  {Aird}, J. 2011, \apj, 743, 46

\bibitem[{Cole {et~al.}(2000)Cole, Lacey, Baugh, \&
  Frenk}]{cole:durham.sam.initial}
Cole, S., Lacey, C.~G., Baugh, C.~M., \& Frenk, C.~S. 2000, \mnras, 319, 168

\bibitem[{{Dav{\'e}} {et~al.}(2011){Dav{\'e}}, {Oppenheimer}, \&
  {Finlator}}]{dave:2011.mf.vs.z.winds}
{Dav{\'e}}, R., {Oppenheimer}, B.~D., \& {Finlator}, K. 2011, \mnras, 415, 11

\bibitem[{{Dutton} \& {van den Bosch}(2009)}]{dutton:2009.wind.scalings.needed}
{Dutton}, A.~A., \& {van den Bosch}, F.~C. 2009, \mnras, 396, 141

\bibitem[{{Erb} {et~al.}(2006){Erb}, {Shapley}, {Pettini}, {Steidel}, {Reddy},
  \& {Adelberger}}]{erb:lbg.metallicity-winds}
{Erb}, D.~K., {Shapley}, A.~E., {Pettini}, M., {Steidel}, C.~C., {Reddy},
  N.~A., \& {Adelberger}, K.~L. 2006, \apj, 644, 813

\bibitem[{{Evans} {et~al.}(2009)}]{evans:2009.sf.efficiencies.lifetimes}
{Evans}, N.~J., {et~al.} 2009, \apjs, 181, 321

\bibitem[{{Evans}(1999)}]{evans:1999.sf.gmc.review}
{Evans}, II, N.~J. 1999, \araa, 37, 311

\bibitem[{{Genel} {et~al.}(2012)}]{genel10}
{Genel}, S., {et~al.} 2012, \apj, 745, 11

\bibitem[{{Governato} {et~al.}(2007)}]{governato:disk.formation}
{Governato}, F., {et~al.} 2007, \mnras, 374, 1479

\bibitem[{{Governato} {et~al.}(2010)}]{governato:2010.dwarf.gal.form}
---. 2010, \nat, 463, 203

\bibitem[{{Guo} {et~al.}(2010){Guo}, {White}, {Li}, \&
  {Boylan-Kolchin}}]{guo:2010.hod.constraints}
{Guo}, Q., {White}, S., {Li}, C., \& {Boylan-Kolchin}, M. 2010, \mnras, 404,
  1111

\bibitem[{{Hall} {et~al.}(2011){Hall}, {Courteau}, {Dutton}, {McDonald}, \&
  {Zhu}}]{hall:2011.disk.scalings}
{Hall}, M., {Courteau}, S., {Dutton}, A.~A., {McDonald}, M., \& {Zhu}, Y. 2011,
  \mnras, in press [arXiv:1111.5009]

\bibitem[{{Heckman} {et~al.}(2000){Heckman}, {Lehnert}, {Strickland}, \&
  {Armus}}]{heckman:superwind.abs.kinematics}
{Heckman}, T.~M., {Lehnert}, M.~D., {Strickland}, D.~K., \& {Armus}, L. 2000,
  \apjs, 129, 493

\bibitem[{{Hernquist}(1990)}]{hernquist:profile}
{Hernquist}, L. 1990, \apj, 356, 359

\bibitem[{{Hopkins} {et~al.}(2005){Hopkins}, {Hernquist}, {Martini}, {Cox},
  {Robertson}, {Di Matteo}, \& {Springel}}]{hopkins:lifetimes.letter}
{Hopkins}, P.~F., {Hernquist}, L., {Martini}, P., {Cox}, T.~J., {Robertson},
  B., {Di Matteo}, T., \& {Springel}, V. 2005, \apjl, 625, L71

\bibitem[{{Hopkins} {et~al.}(2011){Hopkins}, {Quataert}, \&
  {Murray}}]{hopkins:rad.pressure.sf.fb}
{Hopkins}, P.~F., {Quataert}, E., \& {Murray}, N. 2011, \mnras, 417, 950

\bibitem[{{Hopkins} {et~al.}(2012){Hopkins}, {Quataert}, \&
  {Murray}}]{hopkins:fb.ism.prop}
---. 2012, \mnras, 421, 3488

\bibitem[{{Katz} {et~al.}(1996){Katz}, {Weinberg}, \&
  {Hernquist}}]{katz:treesph}
{Katz}, N., {Weinberg}, D.~H., \& {Hernquist}, L. 1996, \apjs, 105, 19

\bibitem[{{Kennicutt}(1998)}]{kennicutt98}
{Kennicutt}, Jr., R.~C. 1998, \apj, 498, 541

\bibitem[{{Keres} {et~al.}(2011){Keres}, {Vogelsberger}, {Springel}, \&
  {Hernquist}}]{keres:2011.aas.arepo}
{Keres}, D., {Vogelsberger}, M., {Springel}, V., \& {Hernquist}, L. 2011, in
  AAS Meeting Abstracts \#218, 119.02--+

\bibitem[{{Kere{\v s}} {et~al.}(2009{\natexlab{a}}){Kere{\v s}}, {Katz},
  {Dav{\'e}}, {Fardal}, \& {Weinberg}}]{keres:fb.constraints.from.cosmo.sims}
{Kere{\v s}}, D., {Katz}, N., {Dav{\'e}}, R., {Fardal}, M., \& {Weinberg},
  D.~H. 2009{\natexlab{a}}, \mnras, 396, 2332

\bibitem[{{Kere{\v s}} {et~al.}(2009{\natexlab{b}}){Kere{\v s}}, {Katz},
  {Dav{\'e}}, {Fardal}, \& {Weinberg}}]{keres:2009.gal.mfs.nofb}
---. 2009{\natexlab{b}}, \mnras, 396, 2332

\bibitem[{{Kroupa}(2002)}]{kroupa:imf}
{Kroupa}, P. 2002, Science, 295, 82

\bibitem[{{Krumholz} \& {Gnedin}(2011)}]{krumholz:2011.molecular.prescription}
{Krumholz}, M.~R., \& {Gnedin}, N.~Y. 2011, \apj, 729, 36

\bibitem[{{Krumholz} \& {Tan}(2007)}]{krumholz:sf.eff.in.clouds}
{Krumholz}, M.~R., \& {Tan}, J.~C. 2007, \apj, 654, 304

\bibitem[{{Leitherer} {et~al.}(1999)}]{starburst99}
{Leitherer}, C., {et~al.} 1999, \apjs, 123, 3

\bibitem[{{Mannucci} {et~al.}(2006){Mannucci}, {Della Valle}, \&
  {Panagia}}]{mannucci:2006.snIa.rates}
{Mannucci}, F., {Della Valle}, M., \& {Panagia}, N. 2006, \mnras, 370, 773

\bibitem[{{Martin}(1999)}]{martin99:outflow.vs.m}
{Martin}, C.~L. 1999, \apj, 513, 156

\bibitem[{{Martin}(2005)}]{martin05:outflows.in.ulirgs}
---. 2005, \apj, 621, 227

\bibitem[{{Martin}(2006)}]{martin06:outflow.extend.origin}
---. 2006, \apj, 647, 222

\bibitem[{{McGaugh}(2005)}]{mcgaugh:tf}
{McGaugh}, S.~S. 2005, \apj, 632, 859

\bibitem[{{Murray} {et~al.}(2011){Murray}, {M{\'e}nard}, \&
  {Thompson}}]{murray:2011.cluster.wind.launching}
{Murray}, N., {M{\'e}nard}, B., \& {Thompson}, T.~A. 2011, \apj, 735, 66

\bibitem[{{Murray} {et~al.}(2005){Murray}, {Quataert}, \&
  {Thompson}}]{murray:momentum.winds}
{Murray}, N., {Quataert}, E., \& {Thompson}, T.~A. 2005, \apj, 618, 569

\bibitem[{{Nagamine}(2010)}]{nagamine:2010.dwarf.gal.cosmo.review}
{Nagamine}, K. 2010, Advances in Astronomy, 2010

\bibitem[{{Newman} {et~al.}(2012)}]{newman:z2.clump.winds.prep}
{Newman}, S.~F., {et~al.} 2012, \apj, in press, arXiv:1204.4727

\bibitem[{{Noeske} {et~al.}(2007)}]{noeske:2007.sfh.part1}
{Noeske}, K.~G., {et~al.} 2007, \apjl, 660, L43

\bibitem[{{Oppenheimer} \& {Dav{\'e}}(2006)}]{oppenheimer:outflow.enrichment}
{Oppenheimer}, B.~D., \& {Dav{\'e}}, R. 2006, \mnras, 373, 1265

\bibitem[{{Oppenheimer} \&
  {Dav{\'e}}(2008)}]{oppenheimer:metal.enrichment.momentum.winds}
---. 2008, \mnras, 387, 577

\bibitem[{{Oppenheimer} {et~al.}(2010){Oppenheimer}, {Dav{\'e}}, {Kere{\v s}},
  {Fardal}, {Katz}, {Kollmeier}, \&
  {Weinberg}}]{oppenheimer:recycled.wind.accretion}
{Oppenheimer}, B.~D., {Dav{\'e}}, R., {Kere{\v s}}, D., {Fardal}, M., {Katz},
  N., {Kollmeier}, J.~A., \& {Weinberg}, D.~H. 2010, \mnras, 406, 2325

\bibitem[{{Pettini} {et~al.}(2003){Pettini}, {Madau}, {Bolte}, {Prochaska},
  {Ellison}, \& {Fan}}]{pettini:2003.igm.metal.evol}
{Pettini}, M., {Madau}, P., {Bolte}, M., {Prochaska}, J.~X., {Ellison}, S.~L.,
  \& {Fan}, X. 2003, \apj, 594, 695

\bibitem[{{Powell} {et~al.}(2011){Powell}, {Slyz}, \&
  {Devriendt}}]{powell:2010.sne.fb.weak.winds}
{Powell}, L.~C., {Slyz}, A., \& {Devriendt}, J. 2011, \mnras, 414, 3671

\bibitem[{{Rupke} {et~al.}(2005){Rupke}, {Veilleux}, \&
  {Sanders}}]{rupke:outflows}
{Rupke}, D.~S., {Veilleux}, S., \& {Sanders}, D.~B. 2005, \apj, 632, 751

\bibitem[{{Sales} {et~al.}(2010){Sales}, {Navarro}, {Schaye}, {Vecchia},
  {Springel}, \& {Booth}}]{sales:2010.cosmo.disks.w.fb}
{Sales}, L.~V., {Navarro}, J.~F., {Schaye}, J., {Vecchia}, C.~D., {Springel},
  V., \& {Booth}, C.~M. 2010, \mnras, 409, 1541

\bibitem[{{Sato} {et~al.}(2009){Sato}, {Martin}, {Noeske}, {Koo}, \&
  {Lotz}}]{sato:2009.ulirg.outflows}
{Sato}, T., {Martin}, C.~L., {Noeske}, K.~G., {Koo}, D.~C., \& {Lotz}, J.~M.
  2009, \apj, 696, 214

\bibitem[{{Somerville} \& {Primack}(1999)}]{somerville99:sam}
{Somerville}, R.~S., \& {Primack}, J.~R. 1999, \mnras, 310, 1087

\bibitem[{{Songaila}(2005)}]{songaila:2005.igm.metal.evol}
{Songaila}, A. 2005, \aj, 130, 1996

\bibitem[{{Springel}(2005)}]{springel:gadget}
{Springel}, V. 2005, \mnras, 364, 1105

\bibitem[{Springel(2010)}]{springel:arepo}
Springel, V. 2010, \mnras, 401, 791

\bibitem[{{Springel} \& {Hernquist}(2003{\natexlab{a}})}]{springel:multiphase}
{Springel}, V., \& {Hernquist}, L. 2003{\natexlab{a}}, \mnras, 339, 289

\bibitem[{{Springel} \& {Hernquist}(2003{\natexlab{b}})}]{springel:lcdm.sfh}
---. 2003{\natexlab{b}}, \mnras, 339, 312

\bibitem[{{Steidel} {et~al.}(2010){Steidel}, {Erb}, {Shapley}, {Pettini},
  {Reddy}, {Bogosavljevi{\'c}}, {Rudie}, \&
  {Rakic}}]{steidel:2010.outflow.kinematics}
{Steidel}, C.~C., {Erb}, D.~K., {Shapley}, A.~E., {Pettini}, M., {Reddy}, N.,
  {Bogosavljevi{\'c}}, M., {Rudie}, G.~C., \& {Rakic}, O. 2010, \apj, 717, 289

\bibitem[{{Thacker} \& {Couchman}(2000)}]{thackercouchman00}
{Thacker}, R.~J., \& {Couchman}, H.~M.~P. 2000, \apj, 545, 728

\bibitem[{{Tremonti} {et~al.}(2004)}]{tremonti:mass.metallicity.relation}
{Tremonti}, C.~A., {et~al.} 2004, \apj, 613, 898

\bibitem[{{Weiner} {et~al.}(2009)}]{weiner:z1.outflows}
{Weiner}, B.~J., {et~al.} 2009, \apj, 692, 187

\bibitem[{{White} \& {Frenk}(1991)}]{white:1991.galform}
{White}, S.~D.~M., \& {Frenk}, C.~S. 1991, \apj, 379, 52

\bibitem[{{Wiersma} {et~al.}(2009){Wiersma}, {Schaye}, \&
  {Smith}}]{wiersma:2009.coolingtables}
{Wiersma}, R.~P.~C., {Schaye}, J., \& {Smith}, B.~D. 2009, \mnras, 393, 99

\bibitem[{{Williams} \& {McKee}(1997)}]{williams:1997.gmc.prop}
{Williams}, J.~P., \& {McKee}, C.~F. 1997, \apj, 476, 166

\bibitem[{{Zuckerman} \& {Evans}(1974)}]{zuckerman:1974.gmc.constraints}
{Zuckerman}, B., \& {Evans}, II, N.~J. 1974, \apjl, 192, L149

\end{thebibliography}


\begin{appendix}

\section{Numerical Tests}
\label{sec:numerical.tests}

Extensive numerical tests of the models used here are presented in 
\paperone\ and \papertwo. We refer to those papers for details, but 
briefly summarize  the results of those numerical tests  for galactic winds.

\begin{figure}
    \centering
    \plotone{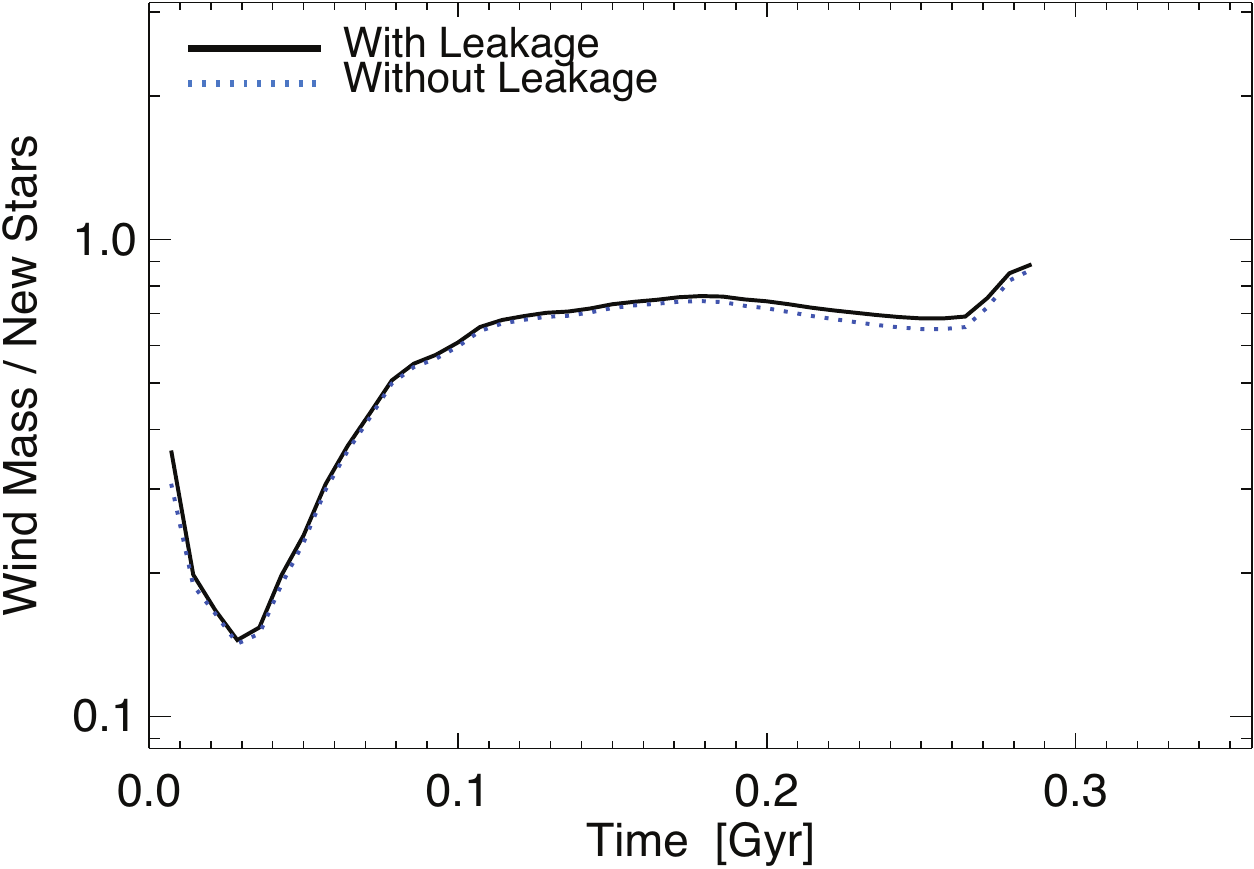}
    \caption{Effects of allowing for photon ``leakage'' -- i.e.\ assuming that the 
    ISM is clumpy and porous on sub-resolution scales. 
    We plot the wind mass-loading versus time (as Figure~\ref{fig:wind.vs.fb}) 
    for an otherwise identical standard HiZ simulation.
    Our standard model (``without leakage'') takes the local column density 
    estimator as-is, and estimates the escape fraction from the spectrum of each star with 
    $f_{\rm esc}=\exp{(-\tau)}$ for each waveband. 
    The ``with leakage'' model assumes that each star is actually surrounded by 
    an (un-resolved) distribution of columns following a power-law 
    distribution (see \paperone\ for details) and uses this to calculate $f_{\rm esc}$.
    The allowance for leakage only makes a difference when the average opacity is large, 
    at which point the contribution (in either case) to the global total luminosity in the relevant band 
    is small. As a result it has a negligible effect on our results.
    \label{fig:leakage}}
\end{figure}

In \paperone, we note that the ISM is probably inhomogeneous below our 
resolution scale, which can allow photons to ``leak out'' from the 
region around their parent star at a rate higher than the nominal 
$\exp(-\tauavg)$ which we use to attenuate the spectrum when calculating 
the radiation pressure effects. This will weaken the short-range radiation pressure, 
but strengthen the long-range radiation pressure forces (since more short-wavelength 
photons escape, to which the ``diffuse'' medium has a higher optical depth). 
In \paperone\ and \papertwo, we show that the ``escape fraction'' (and hence modification 
to these forces) is straightforward to calculate analytically, if we assume some 
{\em a priori} functional form for the sub-grid local density distribution. 
In Figure~\ref{fig:leakage}, we show the consequences for the winds if 
we replace our usual estimator of local attenuation, $f_{\rm esc} = \exp{(-\tauavg)}$, 
with the appropriately modified $f_{\rm esc}$ assuming that 
a gas particle with mean opacity $\tauavg$ really represents a complete 
distribution of opacities with that median but a power-law functional form and dispersion 
of $\approx 0.6\,$dex (chosen to match the median from higher-resolution 
simulations of sub-regions). 
We test this for the HiZ model, since this is the model most sensitive to the 
radiation pressure effects.
Adopting a lognormal distribution gives a 
nearly identical result, for all dispersions $\approx 0 - 2\,$dex (see \paperone). 
In \paperone\ we showed our treatment of the subgrid escape fraction had little effect on the SFR.  Here, we see that it also has no significant effect on the resulting galactic wind. The reason is that for 
low/intermediate $\tauavg$, the escape fractions are large (and hence similar) regardless 
of sub-structure; for high $\tauavg$ the escape fractions can be very different, but they 
are both still small and contribute negligibly to the total coupled momentum.

\begin{figure*}
    \centering
    \plotside{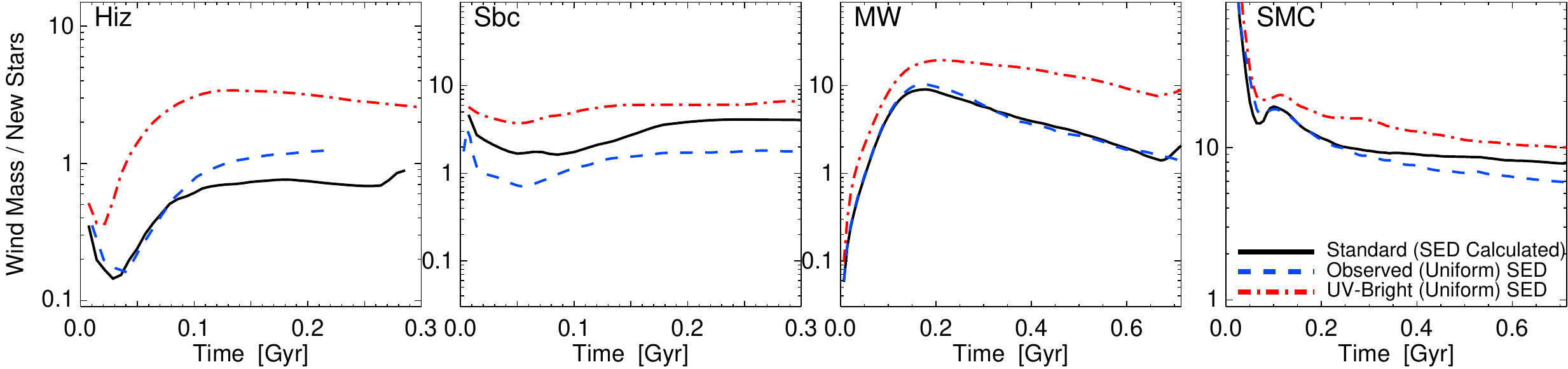}
    \caption{Comparison of our standard long-range radiation pressure 
    model (where the attenuation and SED shape of each star particle is calculated) 
    to one in which we simply force the SED shape to match a specific empirical 
    template. We show the resulting wind mass-loading versus time (as Figure~\ref{fig:wind.vs.fb}) for each of the 
    HiZ/Sbc/MW/SMC models. 
    In each, we compare our standard model (black) in which the SED shapes 
    are self-consistently calculated on-the-fly, to a model where the SED shape 
    ($L_{\rm bol}$ is still determined self-consistently) 
    is forced to a constant fixed value chosen to match the observed mean 
    SED shape for similar observed galaxies to each model (blue). 
    These choices amount to a relative proportion of $L_{\rm bol}$ in the 
    UV/optical/IR bands of $(0.05,\,0.15,\,0.8),\ (0.07,\,0.23,\,0.7),\ 
    (0.1,\,0.4,\,0.5),\ (0.2,\,0.3,\,0.5)$ for the HiZ/Sbc/MW/SMC cases. 
    In each case, the self-consistent and empirically fixed models give SFRs 
    in reasonable agreement; this is because the emergent SEDs calculated 
    with the full models tend to agree well with the typical observed values (see \papertwo). 
    However factor $\lesssim2$ differences 
    do arise, largely because the empirical model does not allow 
    spatial and/or time-dependent variations in the 
    emergent SED (for example, very young stars in the galaxy nucleus tend to be much more obscured 
    than older stars in the galaxy outskirts). 
    We also compare a model with fixed UV/optical/IR fractions, but much higher UV/optical fractions
    than actually observed (red): $(0.2,\,0.3,\,0.5)$ for the HiZ/Sbc/MW cases and $(0.4,\,0.4,\,0.2)$ 
    for the SMC. This artificially boosts the long-range radiation pressure, which leads to a nearly 
    linear corresponding boost in the wind efficiencies.
    \label{fig:radiation.sim.vs.empirical}}
\end{figure*}

In \papertwo, we note that we can isolate the effects of the emergent galaxy SED 
shape on the winds by comparing our full model (in which the 
emergent flux from each star particle in the UV/optical/IR is estimated from 
the stellar age and local attenuation) to one in which we simply set the 
SED shape by hand and force it to be the same 
for all particles. Specifically, we calculate the bolometric luminosity 
and incident force on all gas particles in our standard fashion, but simply 
force the values $L_{\rm UV}/L_{\rm bol}$, 
$L_{\rm Opt}/L_{\rm bol}$, $L_{\rm IR}/L_{\rm bol}$ to equal some chosen values. 

Figure~\ref{fig:radiation.sim.vs.empirical} compares our standard model, 
for each galaxy type, to two choices for this ``enforced SED'' model. 
First, we consider a model in which the SED shape is chosen for each galaxy 
model to match observations of galaxies with similar stellar mass, SFR, and redshift. 
The observations are discussed in \papertwo, but for typical observational 
counterparts to each of our galaxy models, $(f_{\rm UV},\,f_{\rm Opt},\,f_{\rm IR})$ 
is approximately equal to $(0.05,\,0.2,\,0.75)$ for the HiZ, $(0.13,\,0.40,\,0.47)$ for the MW, 
$(0.3,\,0.3,\,0.4)$ for the SMC, 
and $(0.07,0.23,0.7)$ for the Sbc models. There is substantial scatter 
in otherwise similar galaxies and uncertainty in these quantities, but they 
provide an approximate guide. Reassuringly, the resulting wind masses are 
quite similar to our ``standard model,'' so it is likely that the simulations are capturing the key physics 
of radiative acceleration despite lacking a complete on-the-fly radiation hydrodynamics.
This is consistent with the fact that in \papertwo\ we showed that the predicted  SEDs  from 
our radiation pressure model are  similar to those observed.

We also show in Figure~\ref{fig:radiation.sim.vs.empirical} 
the wind mass-loading if we were to assume a uniform SED shape with a 
a significantly higher UV/optical escape fraction. 
For all models, this serves to substantially enhance the 
coupled momentum for material above the disk and so enhances the wind 
mass (and suppresses the SFR, shown in \papertwo). 
This confirms two important points discussed in the text. 
First, on large scales (as opposed to e.g.\ inside GMCs) the 
gas is optically thin in the IR, and so it is the UV/optical photons that 
dominate the long-range radiative acceleration of winds. 
Second, the ``boost'' in the wind mass-loading is approximately linear 
in the UV/optical escape fraction, as predicted in \S~\ref{sec:winds:prediction}. 
This drives the inverse dependence of the wind mass loading 
on gas surface density (at otherwise fixed properties). 

\begin{figure}
    \centering
    \plotone{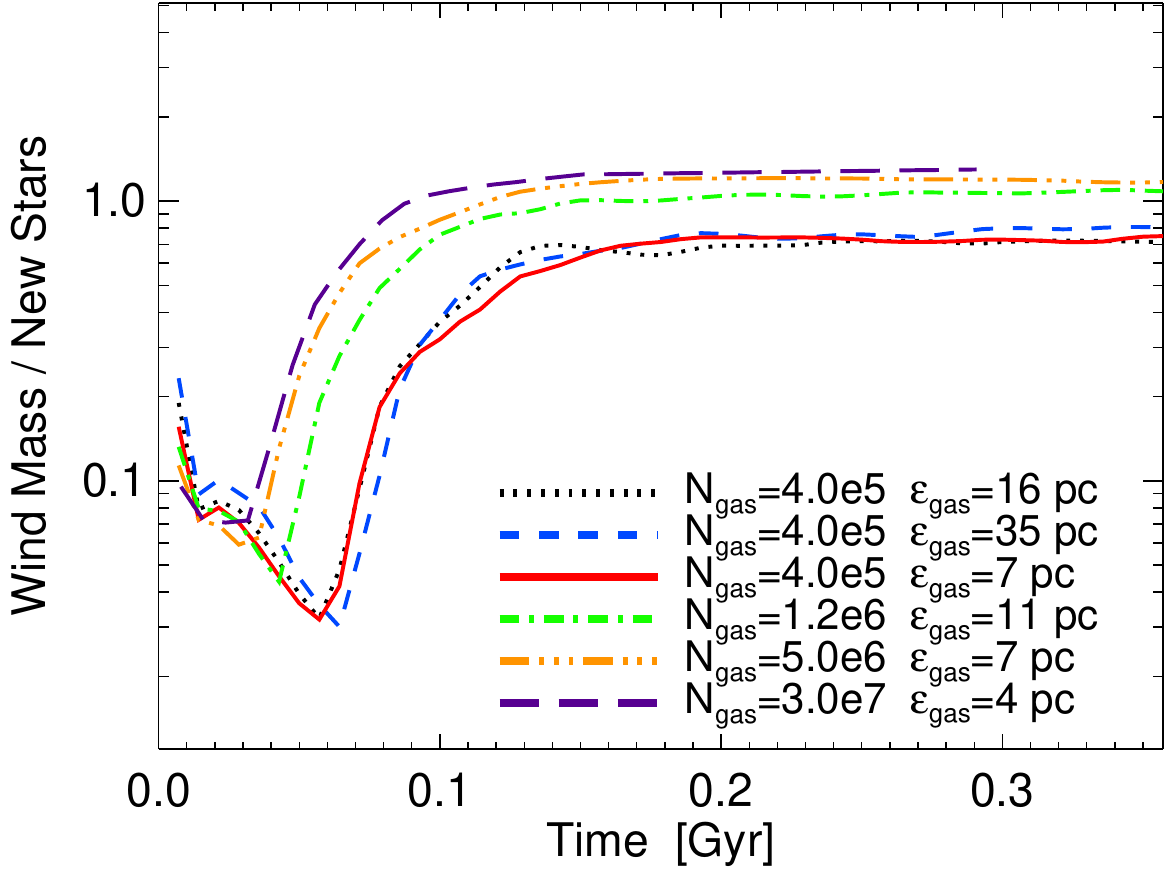}
    \caption{Resolution tests. We plot the wind mass-loading 
    versus time (as Figure~\ref{fig:wind.vs.fb}). 
    We consider a ``standard'' (all feedback enabled) HiZ model at 
    a series in mass and force resolution ($N_{\rm gas}$ is the number of 
    gas particles in the star-forming disk, which 
    determines the mass resolution; $\epsilon_{\rm gas}$ is the 
    force softening). 
    The wind mass tends to increase at higher resolution as venting, mixing, and acceleration of filaments/clumps/streams through a background medium can be more accurately resolved. But once we reach $\sim 10^{6}$ disk particles and $\sim10\,$pc resolution, the winds as well appear to converge (changing by $<10\%$ going to higher resolution). 
    Recall, the Jeans length in these galaxies is several hundred pc 
    (Jeans mass $\gtrsim 10^{8}\,\msun$), so {\em all} of these models formally resolve the 
    Jeans scales. At lower resolution, our prescriptions do not have clear physical meaning. 
    \label{fig:resolution}}
\end{figure}

\vspace{-0.15in}
\subsection{Resolution Tests}

In Figure~\ref{fig:resolution} we consider a basic spatial and mass resolution study of the winds.
We consider an otherwise identical ``standard'' HiZ run, with different particle number (mass resolution) 
and force softening (spatial resolution).\footnote{For 
technical reasons to ensure quantities such as random number generation 
were identical in these runs, this set was run on the same set of processors and so 
does not overlap with our standard HiZ run shown earlier; but the results 
from that run are completely consistent with the standard-resolution case here.}
In \papertwo\ the SFRs for this study are shown -- they converge relatively quickly.
In contrast, the winds show a clear, albeit not dramatic, resolution dependence. 
The fact that the higher-resolution runs converge to equilibrium more rapidly 
is not important -- it is simply a consequence of the fact that most rapidly collapsing spatial scale resolved is on 
a smaller timescale as we go to higher resolution. But going from low to 
intermediate resolution, the equilibrium wind mass loading efficiency approximately doubles. 
At still lower resolution than we show here, the wind mass begins to drop exponentially. This is because 
the simulation can no longer resolve the phases of the ISM and ``holes'' for 
gas to accelerate out along. In addition, at low resolution the columns 
around each star particle (hence extinction of UV/optical light) will be over-estimated, 
since the clumpiness of the ISM is not correctly treated. 
Moreover at low mass resolution it requires very concentrated momentum 
to accelerate even single particles out of the disk -- this is why the results appear to be more 
dependent on the mass resolution ($N_{\rm gas}$), at least in the range we show here, than 
on the force resolution at fixed mass resolution. 
Finally, at low resolution in particular, it is 
well-known that SPH codes with energy and entropy conserving integration schemes 
produce artificial ``surface tension'' forces that suppress the acceleration gas 
streams through a porous or unstable medium. 
Once we reach intermediate resolution, however, the convergence even in the winds becomes quite good. 
From intermediate to standard resolution the wind mass loading changes by only $\sim10-20\%$, 
and from standard to ultra-high resolution it changes by $<5\%$. 

\end{appendix}

\end{document}